\documentclass[preprint]{aastex}

\shorttitle{Telescope Fabra ROA Montsec (TFRM)}
\shortauthors{Fors et al.}

\slugcomment{Submitted to PASP}

\begin{document}

\title{Telescope Fabra ROA Montsec: a new robotic wide-field Baker-Nunn facility}

\author{Octavi Fors\altaffilmark{1,2}}
\email{ofors@am.ub.es}
\author{Jorge N\'u\~nez\altaffilmark{1,2} and Jos\'e Luis Mui\~nos\altaffilmark{3}}

\author{Francisco Javier Montojo\altaffilmark{3} and Roberto Baena-Gall\'e\altaffilmark{1,2}}

\author{Jaime Boloix\altaffilmark{3} and Ricardo Morcillo\altaffilmark{3} and Mar\'\i{}a Teresa Merino\altaffilmark{1,2}}

\author{Elwood C. Downey\altaffilmark{4} and Michael J. Mazur\altaffilmark{5}}

\altaffiltext{1}{Observatori Fabra, Reial Acad\`emia de Ci\`encies i Arts de Barcelona. Rambla dels Estudis, 115. 08002. Barcelona. Spain.}
\altaffiltext{2}{Departament d'Astronomia i Meteorologia and Institut de Ci\`encies del Cosmos (ICC), Universitat de Barcelona
(UB/IEEC). Mart\'i i Franqu\'es 1. 08028. Barcelona. Spain.}
\altaffiltext{3}{Real Instituto y Observatorio de la Armada. Plaza de las Marinas s/n, 11110. San Fernando (C\'adiz), Spain.}
\altaffiltext{4}{Clear Sky Institute Inc., USA}
\altaffiltext{5}{Hygga Innovative Technologies Inc., Norway}

\begin{abstract}

A Baker-Nunn Camera (BNC), originally installed at the Real Instituto y Observatorio de la Armada (ROA) in 1958,
was refurbished and robotized. The new facility, called Telescope Fabra ROA Montsec (TFRM), was installed at the
Observatori Astron\`omic del Montsec (OAdM).

The process of refurbishment is described in detail. Most of the steps of the refurbishment project were
accomplished by purchasing commercial components, which involve little posterior engineering assembling work. The
TFRM is a 0.5m aperture f/0.96 optically modified BNC, which offers a unique combination of instrumental
specifications: fully robotic and remote operation, wide-field of view ($4{\fdg}4{\times}4{\fdg}4$), moderate
limiting magnitude (V${\sim}19.5$\,mag), ability of tracking at arbitrary right ascension ($\alpha$) and
declination ($\delta$) rates, as well as opening and closing CCD shutter at will during an exposure. 

Nearly all kind of image survey programs can benefit from those specifications. Apart from other less time
consuming programs, since the beginning of science TFRM operations we have been conducting two specific
and distinct surveys: super-Earths transiting around M-type dwarfs stars, and geostationary debris in the
context of Space Situational Awareness / Space Surveillance and Tracking (SSA/SST) programs. Preliminary
results for both cases will be shown.

\end{abstract}


\section{INTRODUCTION}

With the launch of Sputnik 1 in the fall of 1957 and other pioneering artificial satellites few months
later, the early Space Age was born. As a solution for optically tracking these new satellites, the
Smithsonian Institution designed and constructed a new kind of telescope: the Baker-Nunn Camera
(BNC)~\citep{1957S&T....16..108H}. The effort invested in such challenging project yielded a cutting edge
prototype, both in terms of technology and optics specifications in that epoch. As a result of those
outstanding specifications, the BNC was able to achieve satellite positional measurements with a typical
accuracy of ${\sim}2{\arcsec}$ for one single station. It is in the context of the International
Geophysical Year, that these measurements allowed to determine for the first time important geophysical
quantities, such as the upper atmosphere drag in satellites orbits, the Earth flattening, the radial
distribution of Earth mass, etc. 

\begin{figure}
\centering 
\includegraphics[angle=0,scale=0.39]{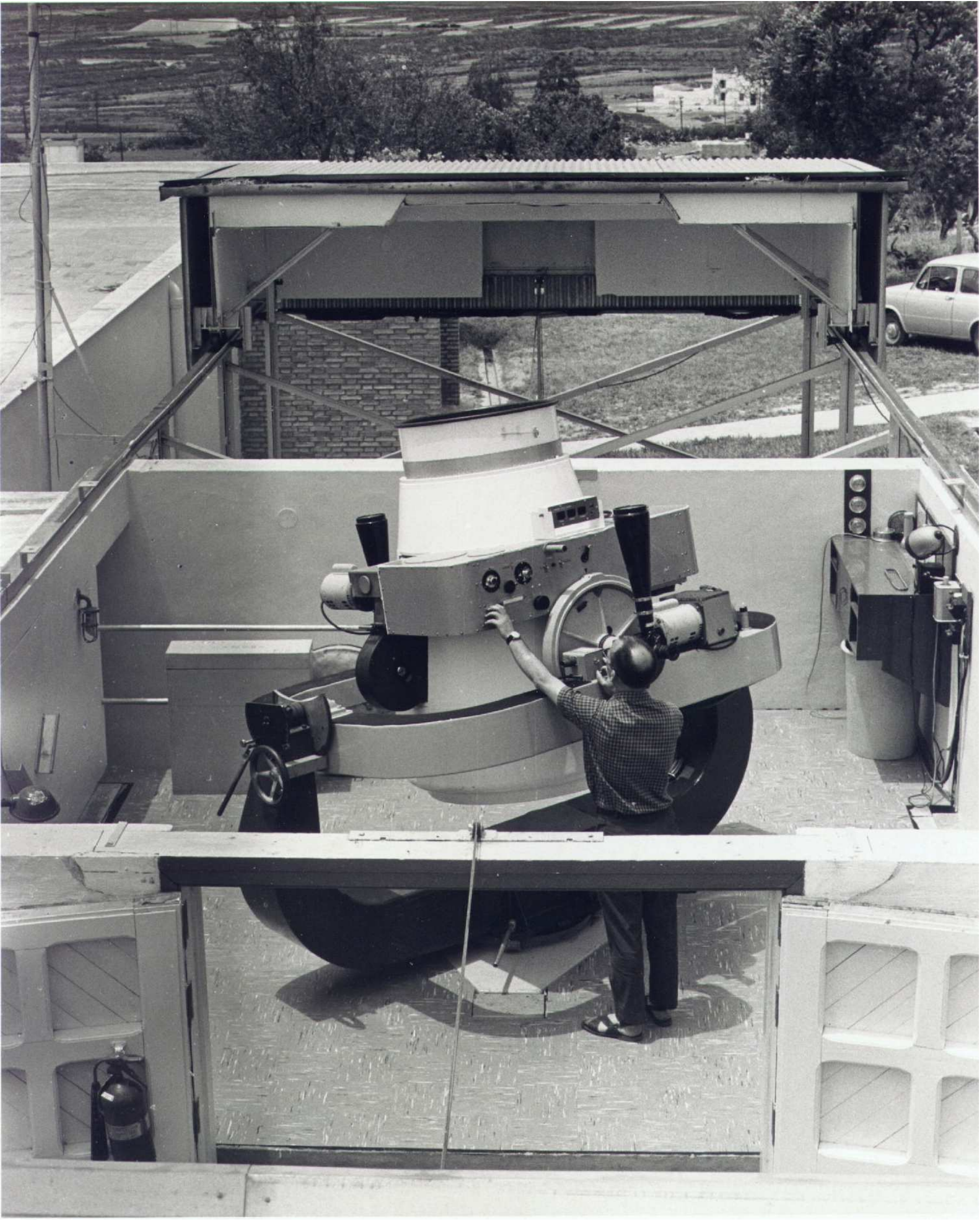}
\caption{Baker-Nunn Camera at ROA in 1958.}
\label{fig:oldbnc}
\end{figure}

In order to maximize the satellites coverage and minimize the positional measurement error, a family of 21
BNCs were manufactured in two releases and placed all over the world spanning in longitude. In 1958 one of
them was installed at ROA (see Fig.~\ref{fig:oldbnc}), in San Fernando (C\'adiz), southern Spain.

\placefigure{fig:oldbnc}

With the upcoming of new satellite tracking technologies on the early 80s, a new kind of facilities called
Ground Based-Electro-Optical Deep Space Surveillance (GEODSS)~\citep{1981MiElC...7...47J} were designed,
manufactured and installed, so that the BNC program became obsolete and was cancelled. The BNC in San
Fernando was donated to ROA, where it was maintained inactive but in excellent state of conservation.

We report here on the refurbishment process of the BNC at ROA, renamed as Telescope Fabra ROA Montsec (TFRM). The
new telescope designation stands for the two partner institutions of the consortium: Reial Acad\`emia de
Ci\`encies i Arts de Barcelona (RACAB) - Observatori Fabra, and Real Instituto y Observatorio de la Armada (ROA),
as well as the observing site: Observatori Astron\`omic del Montsec (OAdM). In summary, the differences between
the original BNC at ROA and the TFRM are: a new motorized equatorial mount, the substitution of the photographic
film with a CCD as a detector, the addition of corrective optics to flatten the CCD field of view (FoV), and the
control software which commands every device of the observatory and formalizes the robotic concept by scheduling
the observing tasks to be executed every night.

In \S~\ref{refurbishment_process} we present in detail the refurbishment process step-by-step with a description
of the specifications of the original BNC at ROA given in \S~\ref{pre_refurbished}, the specifications of
refurbished TFRM in \S~\ref{post_refurbished}, and the observing site in \S~\ref{site}. In \S~\ref{reproducibility},
the reproducibility of the refurbishment process for other BNCs is evaluated.

In \S~\ref{surveying_programs} we show the observational capabilities of the TFRM. In particular, some
demostrative results of the system performance are given in the context of an exoplanet survey and a space
debris survey. Other programs such as surveying the Space Situational Awareness/NEO-segment (SSA/NEO), or
monitoring GRBs, $\gamma$-ray binaries, AGNs, blazars, etc. are also discussed.

In \S~\ref{discussion} we summarize the refurbishment project and TFRM performance after one year of science
operations. We also discuss the {\it know-how} accumulated during the process, and lessons learnt in view of
improving reproducibility for other inactive BNCs.

\section{REFURBISHMENT PROCESS}
\label{refurbishment_process}

The Telescope Fabra ROA Montsec (TFRM) is a consortium that was created to develop and operate a refurbished BNC
(see \url{http://www.am.ub.edu/bnc} for updated details). 

An extensive refurbishment project was conducted using the BNC originally installed at ROA, which successfully
culminated in the TFRM, a wide-field CCD facility with remote and robotic capabilities. In several aspects, our
refurbishment project learnt from the previous experience of the Automated Patrol
Telescope~\citep{1992PASAu..10...74C}, the Phoenix BNC~\citep{2002SPIE.4836..119L}, and the Rothney Astrophysical
Observatory (RAO) BNC~\citep{mazur2005}. We placed special emphasis on the last one, as two of us (RB-G and MTM)
performed a number of research stays at RAO and one of us (MJM), participated in the refurbishment of the RAO BNC.
This allowed us to be more innovative with a number of parts of the project (see \S~\ref{post_refurbished}), and
sped up the learning curve of those parts which were identical. In addition, another refurbishment project for the
Indian BNC, called ARIES~\citep{2005BASI...33..414G}, was planned.

However, it is worth noting that, as far as we know, none of the refurbished BNCs have the ability of
being robotically and/or remotely commanded. Among others, this is the most important difference between
those and our BNC.

\subsection{Specifications of pre-refurbished BNC}
\label{pre_refurbished}

The Baker-Nunn Camera is named after the two pioneering engineers responsible for its optical and mechanical
design, Dr. James G. Baker and Joseph Nunn, respectively. 

\begin{figure}
\centering 
\includegraphics[angle=90,scale=0.70]{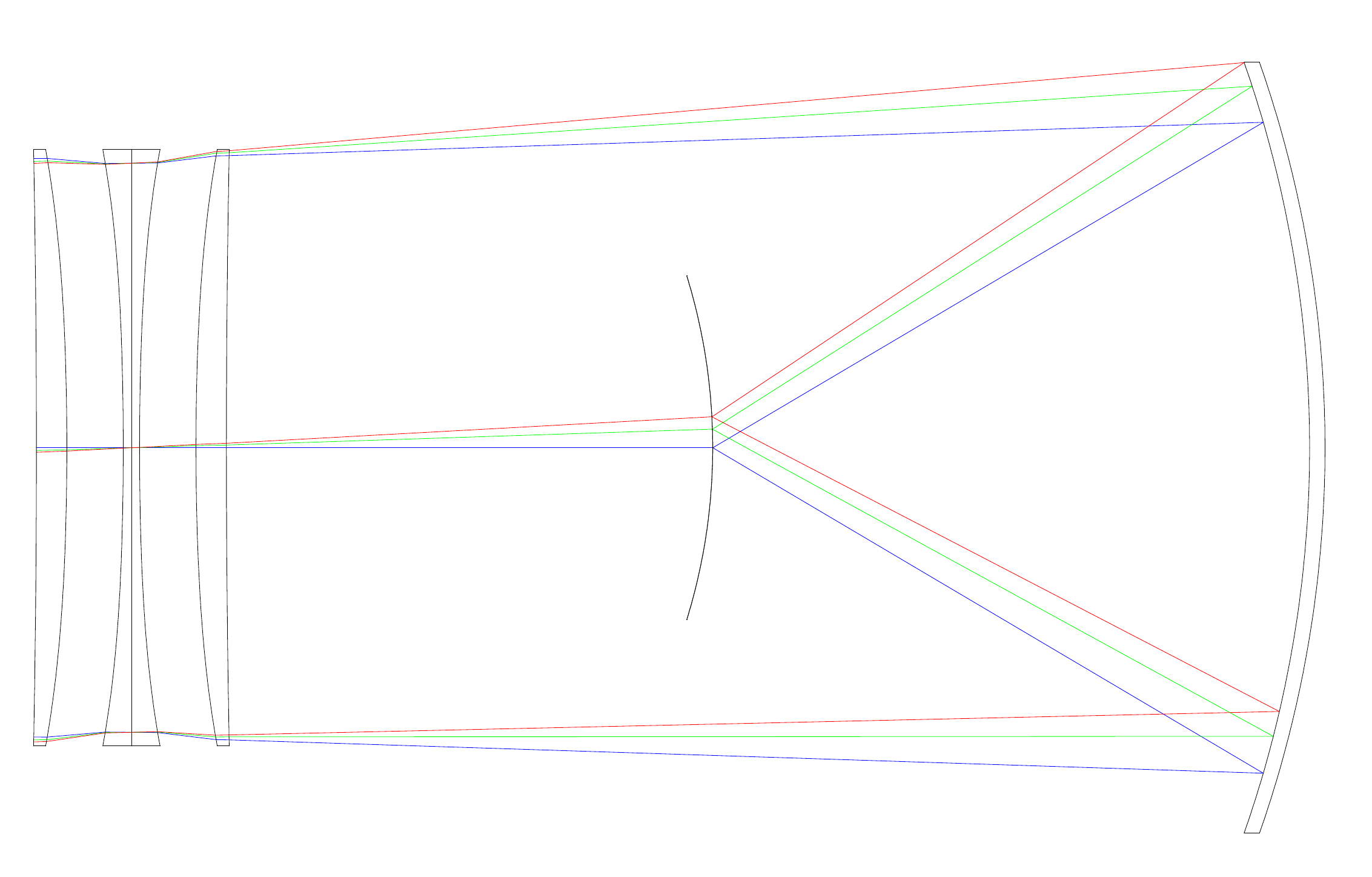}
\caption{Layout of original Baker-Nunn Camera design as defined in Table~\ref{tab:optics_preref}. Axial rays are blocked
by curved focal surface.}
\label{fig:prelayout}
\end{figure}

Optically, the BNC was designed as an f/1 system with 0.5\,m three-element lenses corrector cell, and a 0.78\,m diameter
primary spherical mirror (i.e. modified Schmidt telescope). The two outside surfaces of corrector cell are spherical,
while the other four inner ones are aspherical. The four aspherics are not different each other, but identical in pairs.
The main difference between the BNC and a classical Schmidt system is that the inner aspherical surfaces have more
refractive power, which is necessary for a system as fast as f/1. This, however, adds aberrations that must be accounted
for. To correct this chromatic aberration Dr. Baker made use of a combination of exotic glasses: Schott KzFS-2 and Schott
SK-14 for the outer and inner elements, respectively.  And, as with a Schmidt camera, the focal surface is non-planar.
Along this {\it near} spherical focal surface ran a 55\,mm wide Cinemascope photographic film providing a roughly
30{\degr}x5{\degr} FoV. The optical prescription of the original BNC can be found in Table~\ref{tab:optics_preref}, as
described in~\citet{1962baker}. A drawing of the optical layout for such prescription is shown in
Fig.~\ref{fig:prelayout}.

\placefigure{fig:prelayout}

Mechanically, the BNC was designed to sit on a triangular base. As seen in Fig.~\ref{fig:oldbnc}, mounted on that
base there was a 360{\degr} rotating fork. A gimbal ring was mounted over this fork, so that it could rotate
${\pm}80${\degr} in elevation. The optical tube assembly was mounted on the center of the gimbal ring. The tube
could be motor driven ${\pm70}${\degr} in a 2nd elevation axis that is 90{\degr} to the 1st elevation
axis.
Although complex in design, this balanced alt-alt-az mount could be positioned to
track across any angular direction in the sky - a necessity for early satellite tracking programs.

\begin{deluxetable}{cccccc}
\tabletypesize{\small}
 \tablecaption{Optical design for the pre-refurbished Baker-Nunn Camera.\label{tab:optics_preref}}
 \tablehead{
   \colhead{Surface} & \colhead{Comment} & \colhead{Curvature radius (mm)} &
	 \colhead{Thickness (mm)} & \colhead{Diameter (mm)} & \colhead{Glass}
 }
\startdata
 1&Corrector 1   &-13754.100&      26.54&508&KzFS-2\\
 2&              &-2589.784 &  49.2 &508&\\
 3&Corrector 2   &-2988.818 &  14.25&508&SK-14\\
 4&              &+2988.818 &  49.2 &508&\\
 5&Corrector 3   &+2589.784 &  26.54&508&KzFS-2\\
 6&              &+13754.100&  944.8 &508&\\
 7&Primary mirror&-1016.167 &-520.65&760&Mirror\\
 8&Focus         &-508.001  &&50\tablenotemark{a}&Film\\
\enddata
\tablenotetext{a}{50mm is the width of the film. However, the field is corrected over a 300\,mm diameter surface to
allow for imaging with a 50\,mm${\times}$300\,mm strip of film.}
\end{deluxetable}

\placetable{tab:optics_preref}

When in observing mode, the film was supplied from a large film canister attached to the telescope, stretched over
the focal surface, exposed and then reeled into a take-up canister on the opposite side of the telescope. Two
synchronized, rotating shutters allowed for the trailed images to be {\it chopped} and timestamped with data
provided from the BNC signal clock.

\begin{figure}
\centering 
\includegraphics[angle=0,scale=0.47]{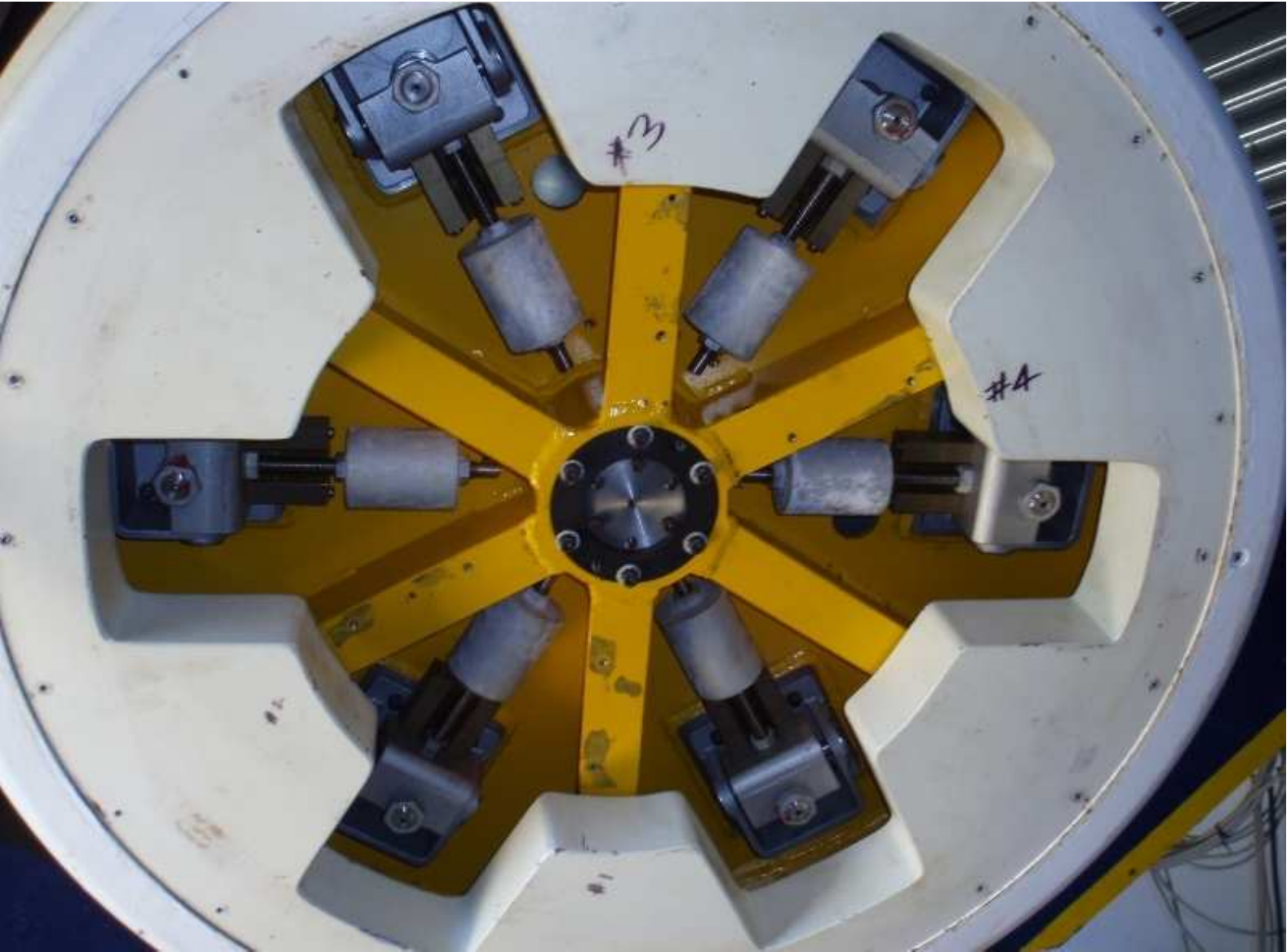}
\caption{Mirror cell with six cylindrical counterweights and a 'floating' mirror which is coupled to the focal
surface through 3 Invar rods. When the pointing or the temperature of the telescope change, the BNC maintains focus
throughout its large FoV.}
\label{fig:mirror_active}
\end{figure}

The primary mirror was suspended inside its cell following an innovative design, which was pioneering before the
upcoming of active optics in the 80s. As seen in Fig.~\ref{fig:mirror_active}, the mirror cell employed a series
of cylindrical counterweights and a 'floating' mirror which was coupled to the focal surface through 3 Invar rods.
These features helped the BNC maintain focus throughout its entire pointing range and over a large range of
temperatures.

\placefigure{fig:mirror_active}

Perkin \& Elmer Corp. was the contractor and manufacturer for the optical grinding, polishing, and figure
testing. Boller \& Chivens manufactured the mechanical parts, and performed the assembly and final
testings.

As a result of the outstanding optical and mechanical designs and the excellent manufacturing process mentioned
above, an extremely high set of optical and mechanical specifications were met. In particular, it was guaranteed
that 80\% of encircled energy of incoming light from UV to deep red was projected within a 20${\mu}$m spot size
throughout the 30{\degr}x5{\degr} FoV covered by the photographic film.

\subsection{Specifications of post-refurbished BNC}
\label{post_refurbished}

\begin{figure}
\centering 
\includegraphics[angle=0,scale=0.75]{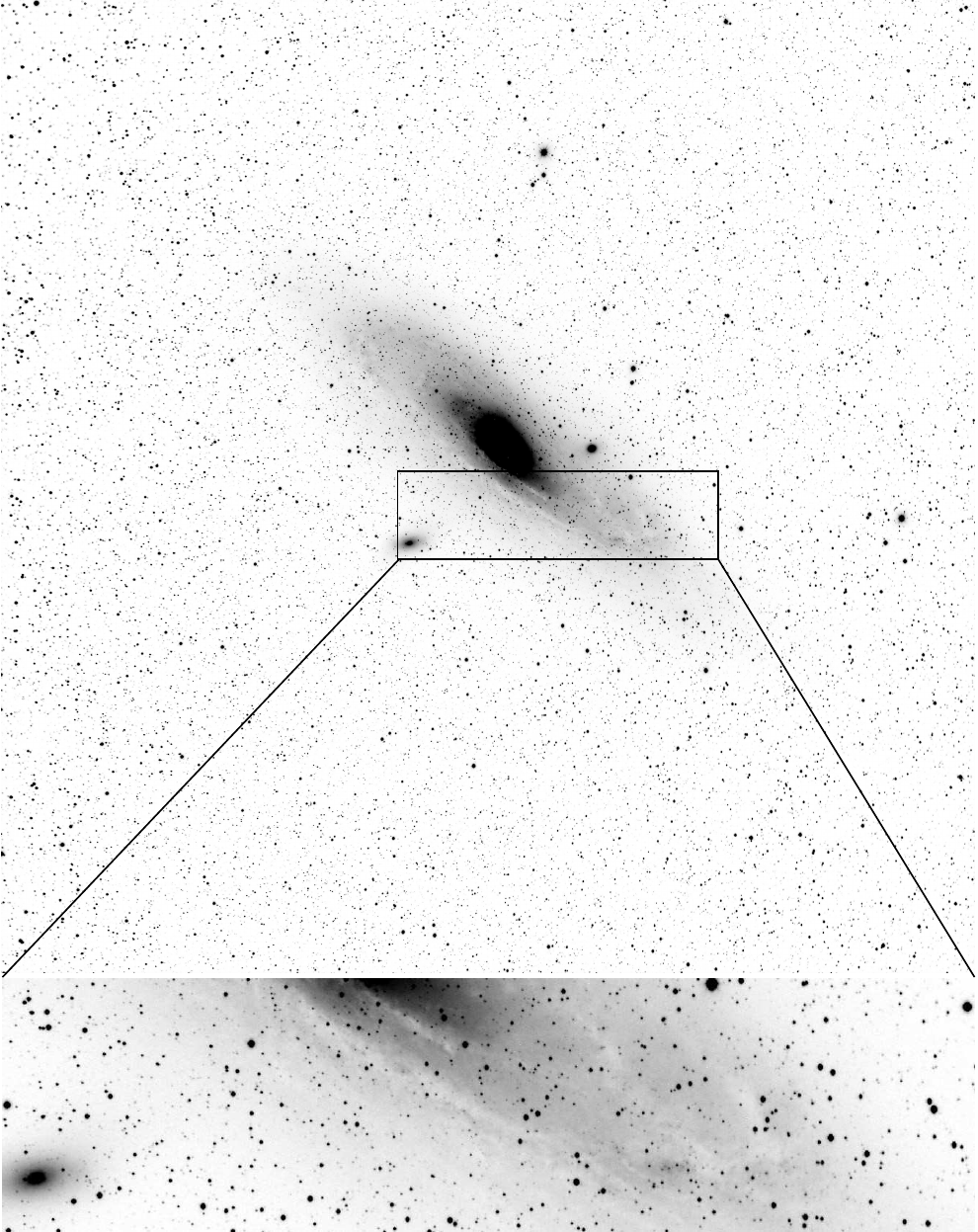}
\caption{{\bf Top:} First 70\,sec exposure technical image of M31 taken at TFRM on September 11, 2010.
Note the huge FoV of the CCD (4.4{\degr}x4.4{\degr}). {\bf Bottom:} Full resolution detail of
spiral arm area.}
\label{fig:firstlightOAdM}
\end{figure}

On September $23^{{\rm rd}}$ of 2009, the (partly) refurbished BNC saw first light at the ROA during a test of its
mount and assessment of the corrector and lens quality. The first images were taken under non-ideal conditions:
urban light polluted skies, unpolished 50cm outermost lens, non-recoated mirror, and uncollimated optics. Despite
these drawbacks, the quality of the first images was very promising. These results were then reconfirmed with the
first technical light at the final observing site (Observatori Astron\`omic del Montsec, OAdM), which was
performed on September $11^{{\rm th}}$ 2010, after mirror recoating, outermost lens repolishing and preliminary
collimation of the optical system. As seen in Fig.~\ref{fig:firstlightOAdM}, the quality is excellent giving the
instrument a great deal of scientific potential.

\placefigure{fig:firstlightOAdM}

\begin{figure*}
\centering 
\includegraphics[angle=0,scale=0.62]{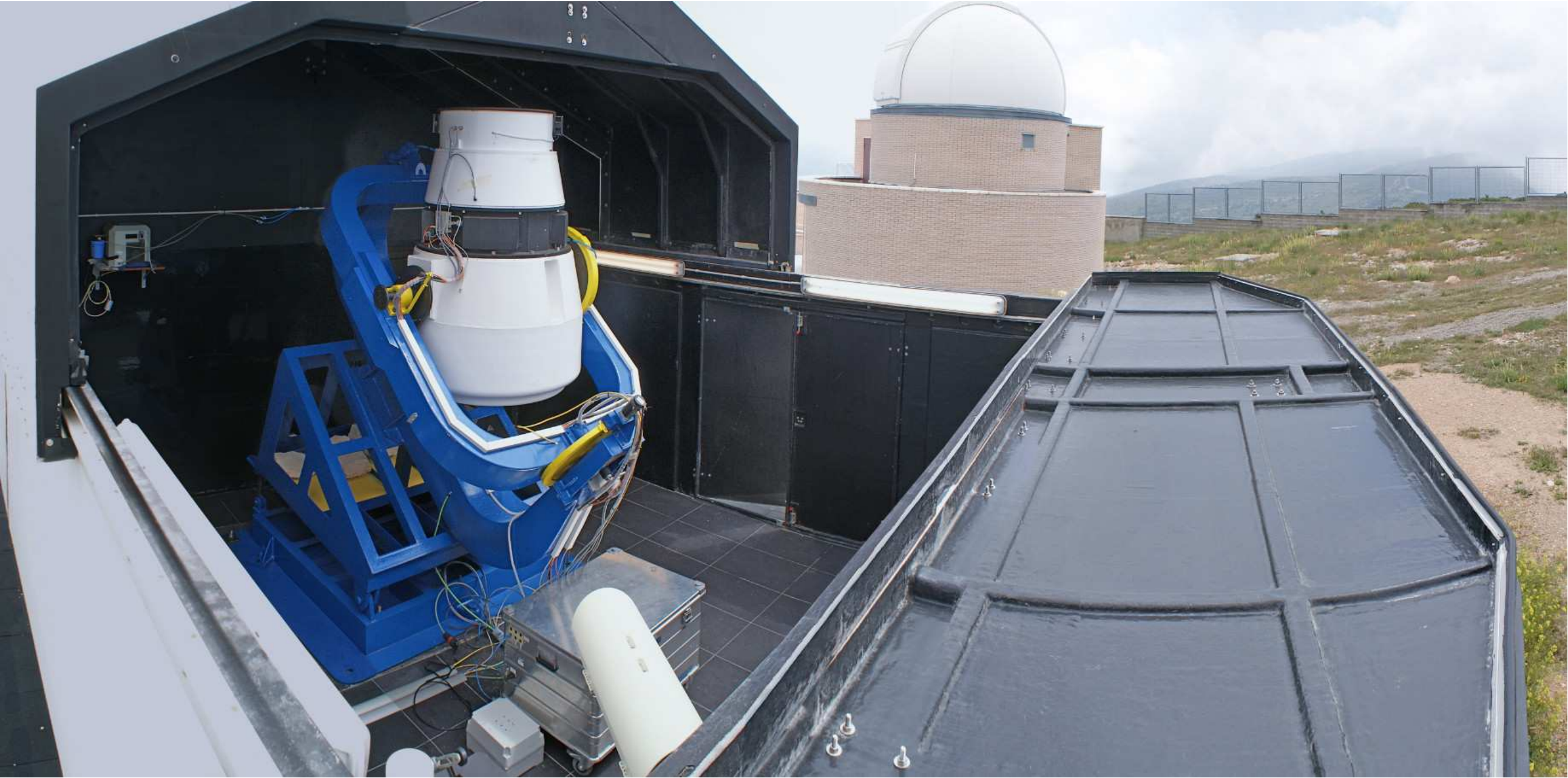}
\caption{{\bf Foreground:} TFRM at OAdM. Sliding roof half open and south gabling wall fully open. Robotic
refurbished BNC inside. {\bf Background:} Joan Or\'o Telescope, also inside OAdM.}
\label{fig:tfrm}
\end{figure*}

A few days later, on 16 Sep 2010, the TFRM was inaugurated~\citep{2010ASPC..430..428F,2010amos.confE..50F}.
Fig.~\ref{fig:tfrm} shows the observatory as it looks now, after conclusion of the refurbishment project. As a
result of this work, the TFRM offers a unique combination of instrumental specifications, namely: a large FoV
($4{\fdg}4{\times}4{\fdg}4$), scale of $3.9{\arcsec}$/pixel, a moderate limiting magnitude (V${\sim}19.5$\,mag),
the capability of tracking at arbitrary $\alpha$ and $\delta$ rates, and the ability to command the CCD camera
shutter at will during an exposure. All-in-all, together with its robotic and remote operations, the TFRM is
strongly suited to conduct observational survey programs (see \S~\ref{surveying_programs}). 

\placefigure{fig:tfrm}

The refurbishment process can be summarized in the following steps:

\subsubsection{Mount modification}

As described in \S~\ref{pre_refurbished}, the original mount design of the BNC had 3 rotational axes (alt-alt-az).

For reasons of reliability and operatibility, it was decided to convert the mount to a 2-axis setup. In deciding
how to proceed, we considered approaches that had been successfully applied during other BNC refurbishments:
\begin{itemize} 
\item[1.] the Automated Patrol Telescope removed the original azimuth base, built a new pier and tilted the yoke
in accordance with the site latitude,
\item[2.] the RAO BNC kept the original azimuth base and tilted this according the latitude for setting
the $\alpha$ axis. In addition, the fork was cut so that original altitude axis is now used as declination axis.
\end{itemize}

The first option is convenient because of its simplicity, specifically for low to moderate latitude sites, where
the installation of a yoke-mounted BNC would be feasible. However, it also imposes restrictions when observing
areas close to the celestial pole. On the other hand, the second alternative requires a major (and more expensive)
transformation of the fork. However, it would be the natural choice for a high latitude site where the first
approach would not be feasible due to the possibility of tube-mount interference. 

Apart from our moderate latitude site, an additional reason led us towards the first approach: the azimuth fork in
early BNCs (like APT and TFRM) was supported by a simple manual bearing assembly. Later BNCs designs (like the
RAO's), however, incorporated a motorized, side-loadable bearing assembly which could be driven while
inclined. Finally, the mount was modified with an adjustable inclination range of ${\pm}5{\degr}$, which spans the
latitudes of the ROA, testing site, and the OAdM, the final observing site.

Other refurbished mount alternatives such as operating in alt-azimuth or buying a new mount were discarded
because of operational difficulties and cost, respectively. 

Once the BNC mount was converted to equatorial mode, both $\alpha$ and $\delta$ axes needed to be motorized such
that they could be commanded from the control computer. As the newly modified $\alpha$ axis lacked a gearing
mechanism, a new 180-teeth gear was machined. A design identical to the original gear in the new $\delta$ axis was
chosen. In addition, a new worm screw for the $\alpha$ axis was machined as a copy of the $\delta$ worm screw,
which was already present in former orbital axis.  

The mount conversion and the manufacture of the $\alpha$ gear was performed by Talleres Yeste S.L. (C\'adiz,
Spain).

Finally, NEMA-234 brushless motors controlled by BearingEngineers AVS digital servo drives were installed on the
$\alpha$ and $\delta$ axes. The telescope closed-loop motion was completed with the following devices:

\begin{itemize}
\item[1.] two 25-bit Heidenhain ECN 225 absolute angle encoders directly installed on each axis shaft and
interfaced via a TCP/IP Heidenhain EIB-741 unit, 
\item[2.] a Pro-Dex PC48 Multi-Axis Motion Controller board plugged on the control computer. A number of
devices of the observatory ($\alpha$ and $\delta$ motors, focus drive, etc.) can be controlled from this
board,
\item[3.] a Meinberg LANTIME M200/GPS time server which allows to synchronize and timestamp UTC time on
whatever observatory device via NTP protocol with an accuracy of ${\pm}0.1$\,ms (limited by CAT6 LAN
latency).
\end{itemize}

\subsubsection{New precise spider and focus system}

A key aspect of the retrofit process is the fabrication of the enclosure and spider
assembly to house the CCD camera and the focus system. All of which sits inside the
Baker-Nunn camera tube. Because of the f/1 nature of the BNC, a ${\pm}10{\mu}$m
repeatability or better focus is required to fully realize the resolution inherent in
the optical system.

\begin{figure*}
\centering 
\includegraphics[angle=0,scale=0.81]{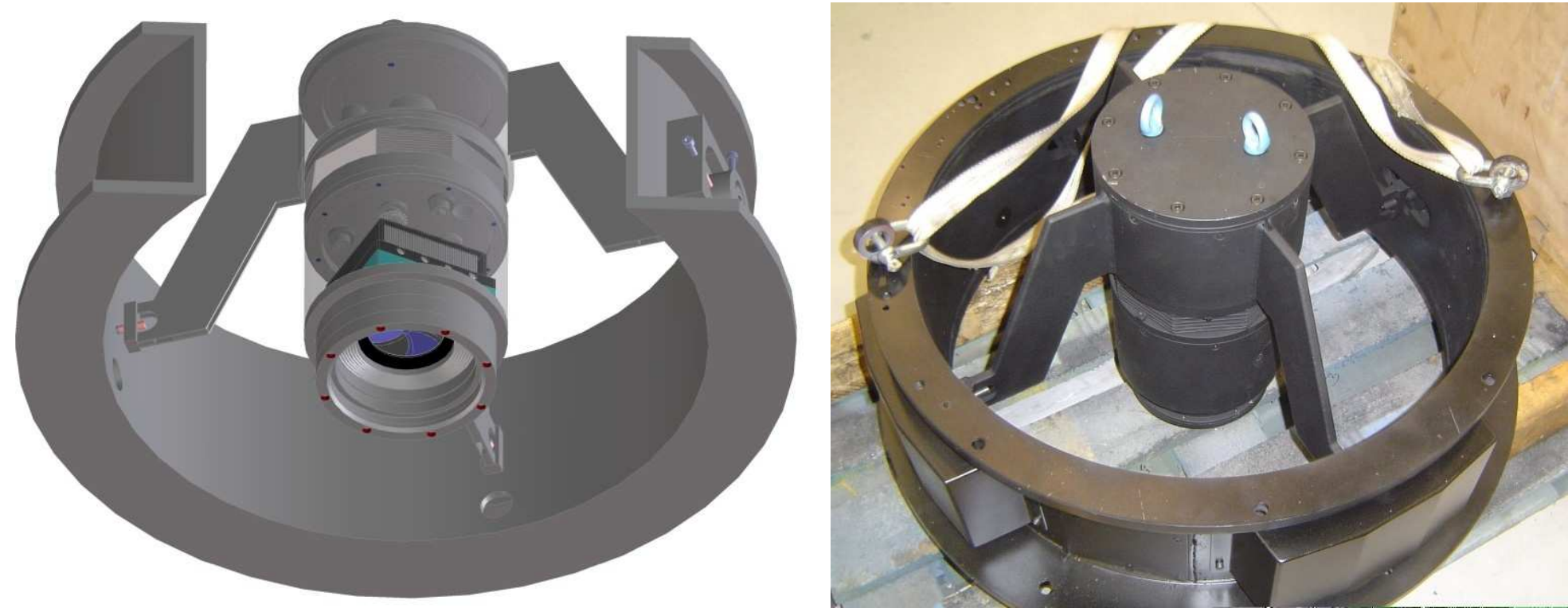}
\caption{{\bf Left:} 3D CAD layout of the design for the spider vanes and CCD focus system. In the central
cylindrical housing, from top to bottom, the focus motor, the CCD camera, its 90\,mm shutter, and the
meniscus hold by eight red-coloured bolts are shown. At the right hand side of the mid tube: tip-tilt
adjuster is shown. {\bf Right:} Finished midtube with spider vanes and CCD housing.}
\label{fig:spider}
\end{figure*}

The solution to these requirements was as follows:

\begin{itemize} 

\item[1.] a preliminary design for a tilt/rotate camera support assembly was provided by MJM based on his
experience with the refurbishment of the RAO BNC.

\item[2.] the final, optimized design was developed and fabricated by Moreno Pujal S.L. (Barcelona, Spain), as can
be seen in Fig.~\ref{fig:spider}. This consists on a steel vane assembly that attaches to a central focus housing.
That is a cylindrical steel shell containing the focus mechanics and motor. The CCD camera housing is attached to
the end of a triple-contact focusing ram. At the bottom of the CCD housing a cell keeps the meniscus lens at the
prescribed distance to the CCD chip. The whole assembly is attached to four spider vanes which are attached to the
TFRM midtube section. Further, precise alignment of the camera with respect to the optical axis of the telescope
is performed using rotational and tip-tilt adjusters installed in the midtube section. These adjusters have a
tip-tilt resolution of $3{\mu}$m, a rotational resolution of 43{\arcsec}, and radial resolution of $63{\mu}$m.

\end{itemize}

\placefigure{fig:spider}

A remarkable feature of the design specified by Moreno Pujal S.L. is the capability of removing the central focus
and CCD housings without touching the spider vanes. This requirement helps ensure optimal aligment of the whole
assembly when the central cylinder needs to be removed for maintainance. Operational experience since Sep 2010
indicates that such a specification has been met: for the four times that the cylindrical housing has been
removed for upgrading purposes, only small variations on the collimation of the system were noted upon reassembly.
This means that relatively few hours have been needed for recollimation after dismantling/reassembling the BNC
camera housing unit.

Another critical specification was that the spider vanes and focus system be as athermal as possible. In other words, with
an f/1 system we could not afford to have focus changes larger than ${\pm}10{\mu}$m due to temperature changes, difference
in telescope attitude, etc. The focus stability after two years of science operations has shown outstanding performance:
not only is the focus adjustment unnecessary during a given night, but we have realized that focus can be maintained
unchanged during months with no loss in faintest object detection.

Finally, the inclusion of two baffles between the meniscus and CCD shutter and the coating of the
internal sides of the CCD housing with an special ultra-black material (MagicBlack, Acktar Inc.) serves
for minimizing the background level and eliminating ghost images due to internal reflections.

\subsubsection{Optics refiguring}
\label{opticsrefiguring}

The original Baker-Nunn 30{\degr}x5{\degr} FoV required a curved focal surface. With the commercial CCD detector
used for this refurbishment, a focal plane was mandatory. This new design required the manufacture of three new
elements: a bi-convex field flattening lens, a meniscus lens, and a plano-plano colour filter. In addition, both
the outermost surface of the 50\,cm corrector cell and the primary mirror had to be repolished and recoated
respectively to get maximum throughput of the system. 

As a result, this corrected design yielded an f/0.96 modified BNC system with a $4{\fdg}4{\times}4{\fdg}4$ (more
than five degrees diameter) FoV which comfortably placed more than 80\% of the ensquared energy within a
$20{\mu}$m spot. At the extreme field point ($3{\fdg}125$), the ensquared energy falls to just over 65\% within
$20{\mu}$m.

To accomplish this, the following general requirements were specified to Malcolm J. MacFarlane, the engineer hired
to perform the optical design:

\begin{itemize}
\item[1.] the focus surface must be flat, which is not the case of the original BNC design.
\item[2.] the geometric distortion of the camera must be less than $10{\mu}$m (0.03\%) at the edge of the
field.
\item[3.] the flat field size must be $6{\fdg}25$ in diameter.
\item[4.] the useful spectral region is 450-1100\,nm.
\end{itemize}

\begin{figure*}
\centering 
\includegraphics[angle=90,scale=0.70]{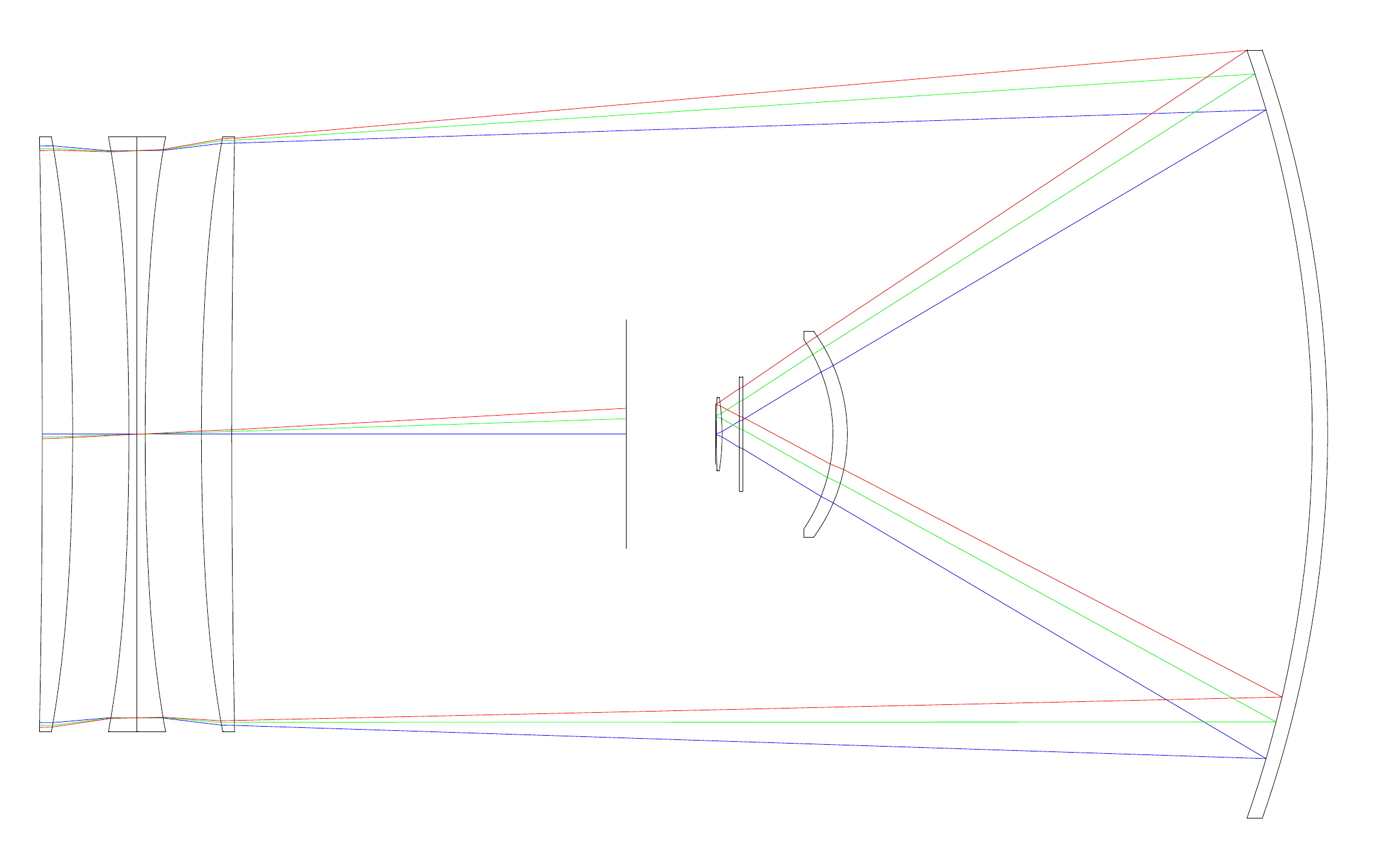}
\caption{Layout of the corrected Baker-Nunn Camera design coinceived for the TFRM as defined in
Table~\ref{tab:optics_postref}. Axial rays are blocked by spider central obstruction.}
\label{fig:postlayout}
\end{figure*}

A final design was accomplished with the specific parameters in Table~\ref{tab:optics_postref}. This design was inspired
by the work which Dr. MacFarlane had already done for the RAO BNC with the addition of a  more stringent requirement on
geometric distorsion. In order to meet the distortion requirement and to obtain similar encircled energy figures as the
original Baker prescription, the use of an unusual material (CaF$_{2}$) for the field flattener and an elliptical surface
on the mensicus corrector were required. A drawing of the optical layout for such corrected prescription is shown in
Fig.~\ref{fig:postlayout}.

\begin{deluxetable}{cccccc}
 \tabletypesize{\small}
 \tablecaption{Optical design for the post-refurbished Baker-Nunn Camera.\label{tab:optics_postref}}
 \tablehead{
   \colhead{Surface} & \colhead{Comment} & \colhead{Curvature (mm)} & 
	 \colhead{Thickness (mm)} & \colhead{Diameter (mm)} & \colhead{Glass}}
\startdata
Object&                                    &$\infty$ &$\infty$ &0       &\\
 1    &Corrector 1                         &-13754.1 &26.54    &508&KzFS-2\\
 2    &                                    &-2589.784&49.2     &508     &\\
 3    &Corrector 2                         &-2988.818&7.125    &508     &SK-14\\
 4    &Stop                                &$\infty$ &7.125    &508     &SK-14\\
 5    &                                    &+2988.818&49.2     &508     &\\
 6    &Corrector 3                         &+2589.784&26.54    &508     &KzFS-2\\
 7    &                                    &+13754.1 &344.8    &508     &\\
 8    &CCD camera shadow                   &$\infty$ &600      &200     &\\
 9    &Primary mirror                      &-1016.167&-406.5262&671.8549\tablenotemark{a}&Mirror\\
10    &Meniscus lens                       &-152.11  &-12.624  &180     &Fused silica\\
11    &Ellipsoidal surface\tablenotemark{b}&-146.887 &-78.87521&166     &\\
12    &Filter                              &$\infty$ &-3.01752 &100     &K5\\
13    &                                    &$\infty$ &-15      &100     &\\
14    &Field flattenner                    &-210.4949&-5.05206 &64      &CaF$_{2}$\\
15    &                                    &+822.5282&-0.65    &64      &\\
Image &Focus                               &$\infty$ &         &52.38919&\\
\enddata
\tablenotetext{a}{Diameter of illuminated circle for a 5${\degr}$ FoV.}
\tablenotetext{b}{Conic constant of ellipsoidal surface=$-0.06049{\pm}0.0005$\,mm.}
\end{deluxetable}

\placetable{tab:optics_postref}

\placefigure{fig:postlayout}

The fact that the field flattener has to be placed as close as 0.65\,mm to the CCD chip and that it is
made of calcium fluoride introduces some restrictions in the cooling rate of the CCD dewar. However, the
flattener was designed thin enough to reach thermal equilibrium without any significant risk of breakage.
In addition, an antireflection coating was applied to both surfaces of the flattener lens.

The meniscus lens was required to correct for the astigmatism introduced by the field flattening lens. To do so,
the optical design placed that element far from the focus plane and outside the CCD camera body. With such a fast
optical system, however, this means that the diameter of the meniscus lens becomes large (180\,mm). In addition to
correcting for astigmatism, the meniscus corrector was also designed to correct for barrel distortion. To
accomplish this, the meniscus lens has deep surfaces - one of which is ellipsoidal. Also, an antireflection
coating was applied to both surfaces of the lens.

Because of the many observational programs to be conducted with the TFRM, the use of a filter is
desirable. Johnson interference filters were early discarded because of the unavoidable chromatic
aberration with the great incidence angle of the f/0.96 beam. Therefore, a coloured glass filter had to be
chosen. Since original BNC optics was not optimized for blue wavelengths and the inferior efficiency of
the CCD in this part of the visible spectrum, a yellow glass filter with a cutoff frequency of 475\,nm
(Schott GG475) was found to be the best choice. Again, antireflection coating was applied to both surfaces
of the filter.

In summary, the redesign layout consisted on the focus surface being flattened by means of adding a positive lens
very close to the CCD and a meniscus lens somewhat further from the focus plane. This latter element provided
correction for the astigmatism introduced by the field flattener. In order to keep the field flattener from
introducing unacceptable aberrations, it was necessary to place it 0.65\,mm from the focal plane.

Furthermore, in order to increase the throughtput of the system, the transparency and reflectivity of the 
outermost 50\,cm surface of the corrector cell and the primary mirror, respectively, had to be improved.

The exterior 50\,cm lens element of the corrector cell is made of KzFS-2, which is a highly hygroscopic glass. As
a result, during the years the BNC was inactive and exposed to ambient humidity, the transparency of this element 
decreased significantly. This decrease in transparency was also observed to occur, to one degree or another, in
other BNCs which did not have in origin a protective plate of the corrector cell. This is the case of the
APT~\citep{1992PASAu..10...74C}.

\begin{figure*}
\centering 
\includegraphics[angle=0,scale=0.78]{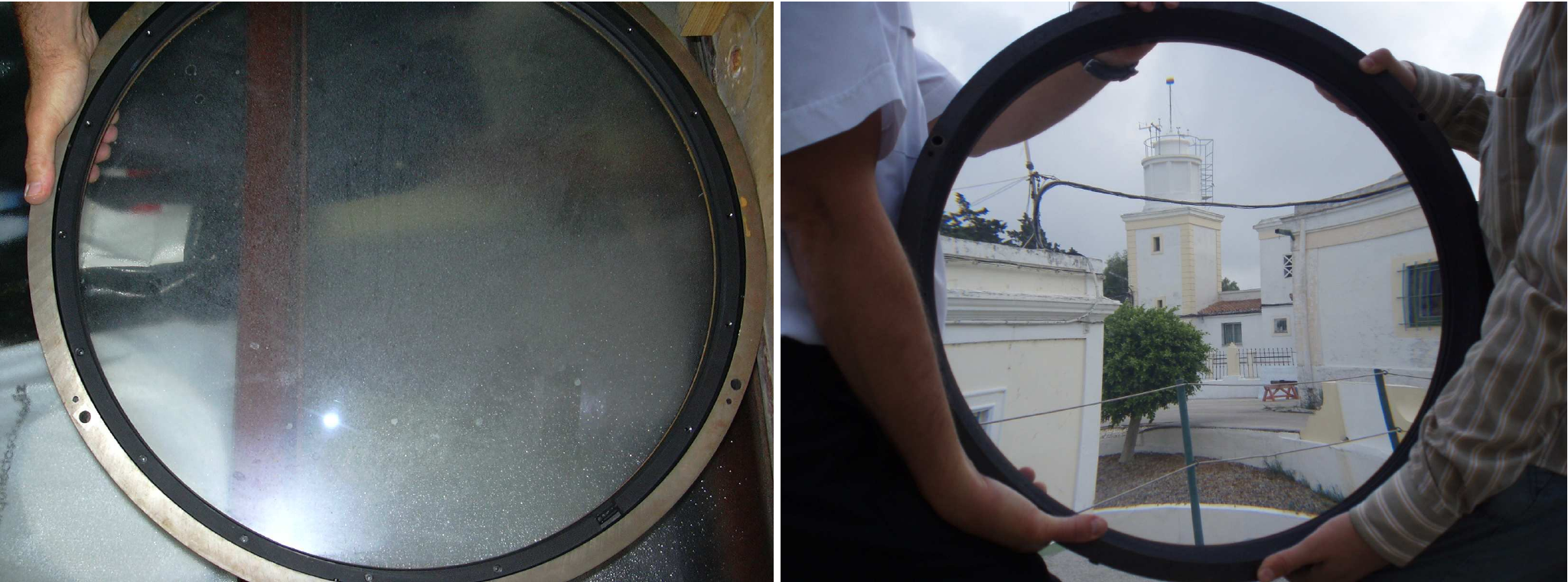}
\caption{Outermost lens before (left) and after (right) repolishing.}
\label{fig:opticslens}
\end{figure*}

A repolishing of the outermost spherical surface was sufficient to restore the original transparency. In addition,
in order to protect the lens from humidity damage in the future, the outermost surface was coated with MgF$_2$
layer. In Fig.~\ref{fig:opticslens} the evident transparency improvement due to the repolishing operation can be
appreciated.

\placefigure{fig:opticslens}

\begin{figure*}
\centering 
\includegraphics[angle=0,scale=0.78]{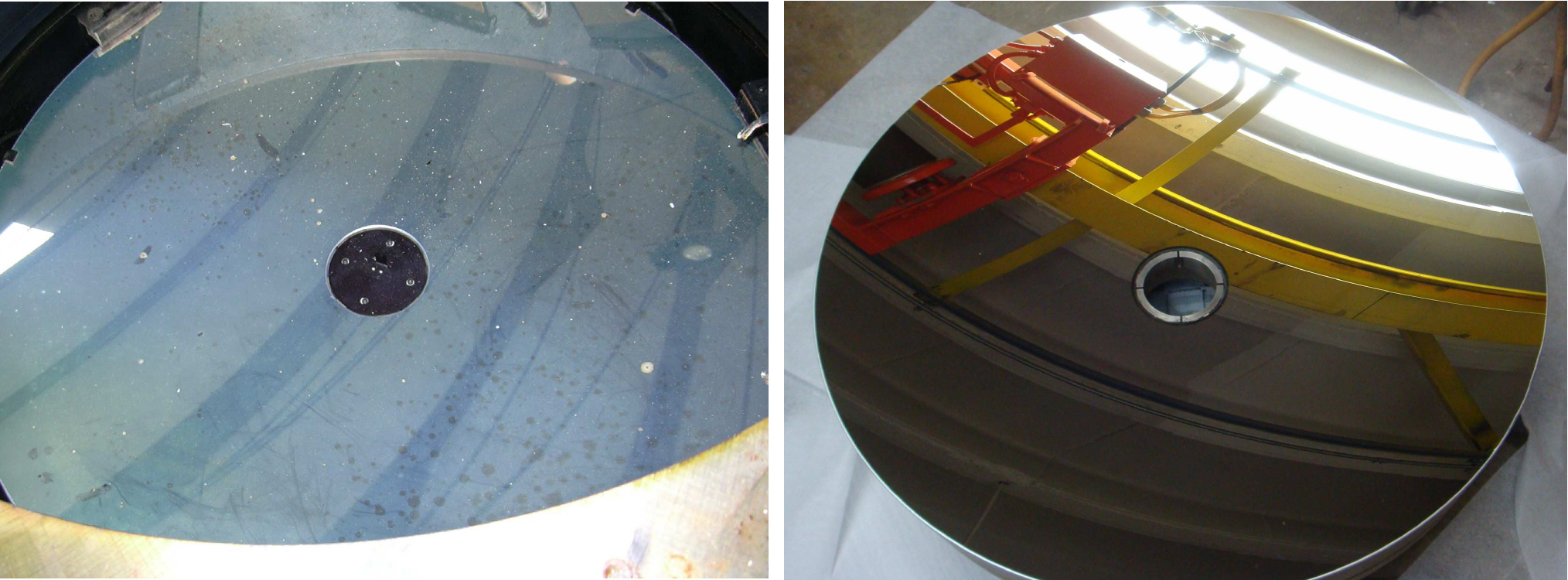}
\caption{Primary mirror before (left) and after (right) recoating.}
\label{fig:opticsmirror}
\end{figure*}

Despite being sealed within the tube, the aluminized surface of the mirror had lost much of its reflectivity.
Because of this, it was necessary to recoat the surface of the mirror. Given the special characteristics of BNC
system (with mirror being difficult to remove from tube), a very durable reflective coating was chosen over other
criteria. The coating Diamond-Brite$^{\rm TM}$ from H.L.Clausing Inc. (Illinois, USA) was chosen for its
durability. In Fig.~\ref{fig:opticsmirror} the increase of mirror reflectivity before and after the recoating
operation is shown.

\placefigure{fig:opticsmirror}

The manufacture of the field flattener lens, the CCD filter, and the repolishing of outermost 50\,cm lens were
performed by Harold Johnson Optical Laboratories Inc (California, USA). The manufacturance of the ellipsoidal
meniscus was conducted by Tucson Optical Research Corporation Inc (Arizona, USA). Mirror recoating was performed
by H.L.Clausing, Inc. 

For the case of lens repolishing and mirror coating their original prescription parameters in Table~\ref{tab:optics_preref}
were not modified. Regarding the field flattener lens, CCD filter and ellipsoidal meniscus, the comparison of the as-built
optics with the theoretical prescription show that all these surfaces were manufactured well within the design tolerances
(see Table~\ref{tab:optics_asbuilt}).

\placetable{tab:optics_asbuilt}

\begin{deluxetable}{ccccc}
\tabletypesize{\small}
 \tablecaption{Comparison of as-built new corrective optical surfaces with theoretical prescription values.\label{tab:optics_asbuilt}}
 \tablehead{
   \colhead{Surface\tablenotemark{a}} & \colhead{Parameter} & \colhead{Prescribed value (mm)} &
	 \colhead{Tolerance (mm)} & \colhead{As-built value (mm)}
 }
\startdata
\sidehead{Meniscus}
 10&Convex curvature                    &-152.1292 &${\pm}$0.25&-152.110  \\
 11&Concave curvature                   &-146.920  &${\pm}$0.25&-146.887  \\
 11&Fringe irregularity\tablenotemark{b}&1         &5          &          \\
 10 \& 11&Axial thickness               &12.700    &${\pm}$0.10&12.624    \\
\sidehead{Filter}
 12 \& 13&Thickness                     &2.99974   &${\pm}$0.10& 3.01752  \\
\sidehead{Field flattener}
 14&Curvature 1                         &-210.4624 &${\pm}$1.0 &-210.49488\\
 15&Curvature 2                         &+823.0108 &${\pm}$1.0 &+822.5282 \\
 14 \& 15&Axial thickness               & 5.05206  &${\pm}$0.10& 5.00126  \\
\enddata
\tablenotetext{a}{Surfaces numbers as noted in Table~\ref{tab:optics_postref}.}
\tablenotetext{b}{Number of fringes on the aspheric surface as visible on a Zygo interferogram over a 2${\arcsec}$ diameter centered area.}
\end{deluxetable}

\begin{figure*}
\centering 
\includegraphics[angle=0,scale=0.69]{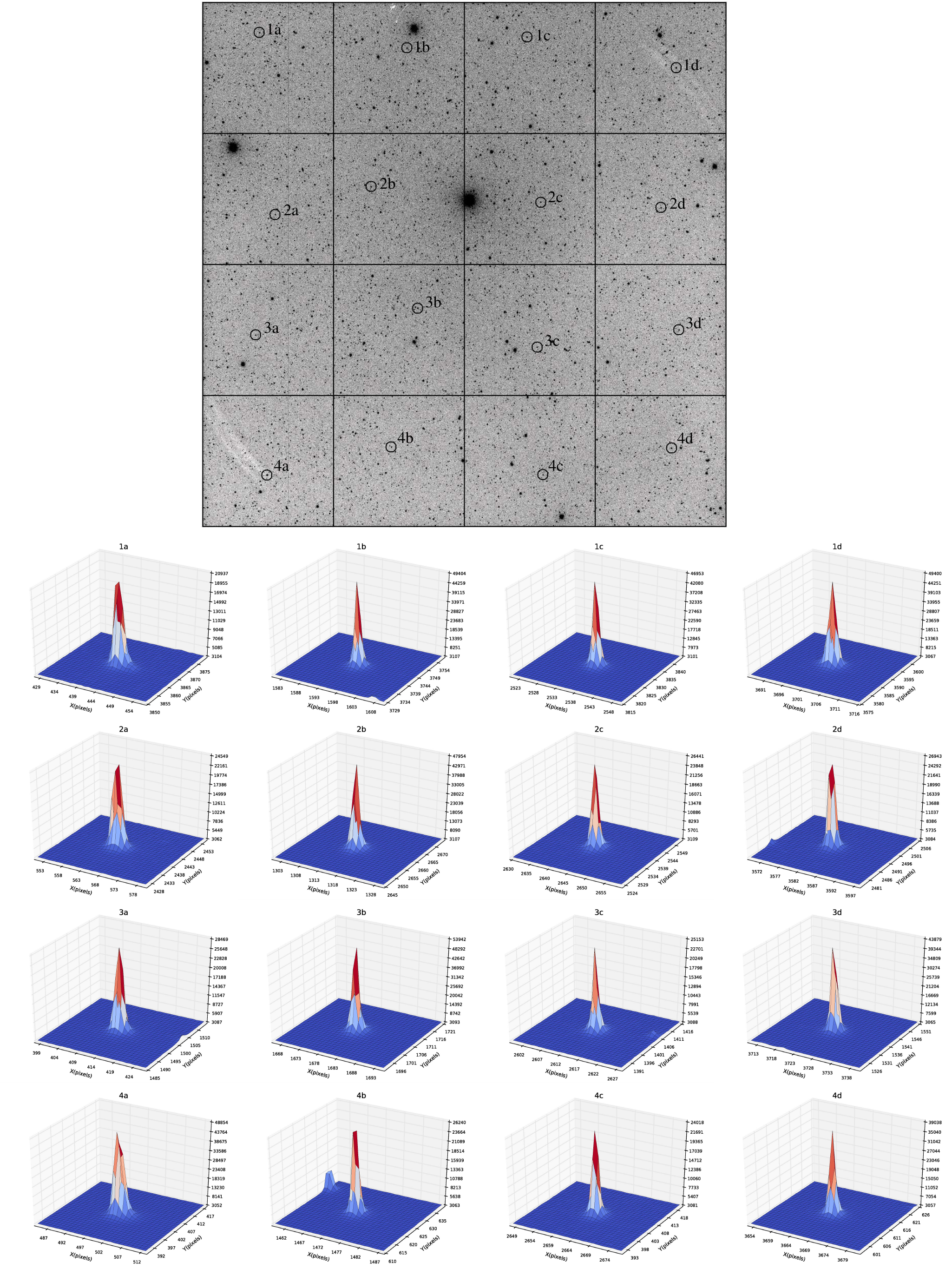}
\caption{A quantitative plot of achieved post-refurbishment point spread function (PSF) as a function of field
angle. Each encircled star on the top image is plotted in a 30${\times}$30pixels 3D surface. From these,
there does not appear to be significant image degradation with increasing field angle.}
\label{fig:PSFangle}
\end{figure*}

Star testing of the optical system, post-refurbishment, shows that the acquired images are, for the most part, free from
obvious aberrations (Fig.~\ref{fig:PSFangle}). Furthermore, it can be seen that image quality (as measured by point spread
function) is consistent with increasing field angles. Note that the image in the figure has not been calibrated (bias, dark
and flatfield) in order to make sure the displayed stellar surfaces are the ones which the optic system project over the
chip. Note the two circular scratches can be seen on the upper right and lower left corners.

\placefigure{fig:PSFangle}

After repolishing and coating with MgF$_2$, the 3-lens corrector was tested to have a 69\% throughput. The meniscus, field
flattener, and filter are all coated with BBAR coatings with a reflectivity of about 1\% (on average) at each surface. So,
the transmission for these three elements totals 94\%. The mirror was coated with Clausing's Diamond Brite$^{\rm TM}$
coating which has a stated reflectivity of 97\%. So, in total, 63\% of the incoming light reaches the chip.

\subsubsection{Custom CCD camera}

\begin{figure}
\centering 
\includegraphics[angle=0,scale=0.47]{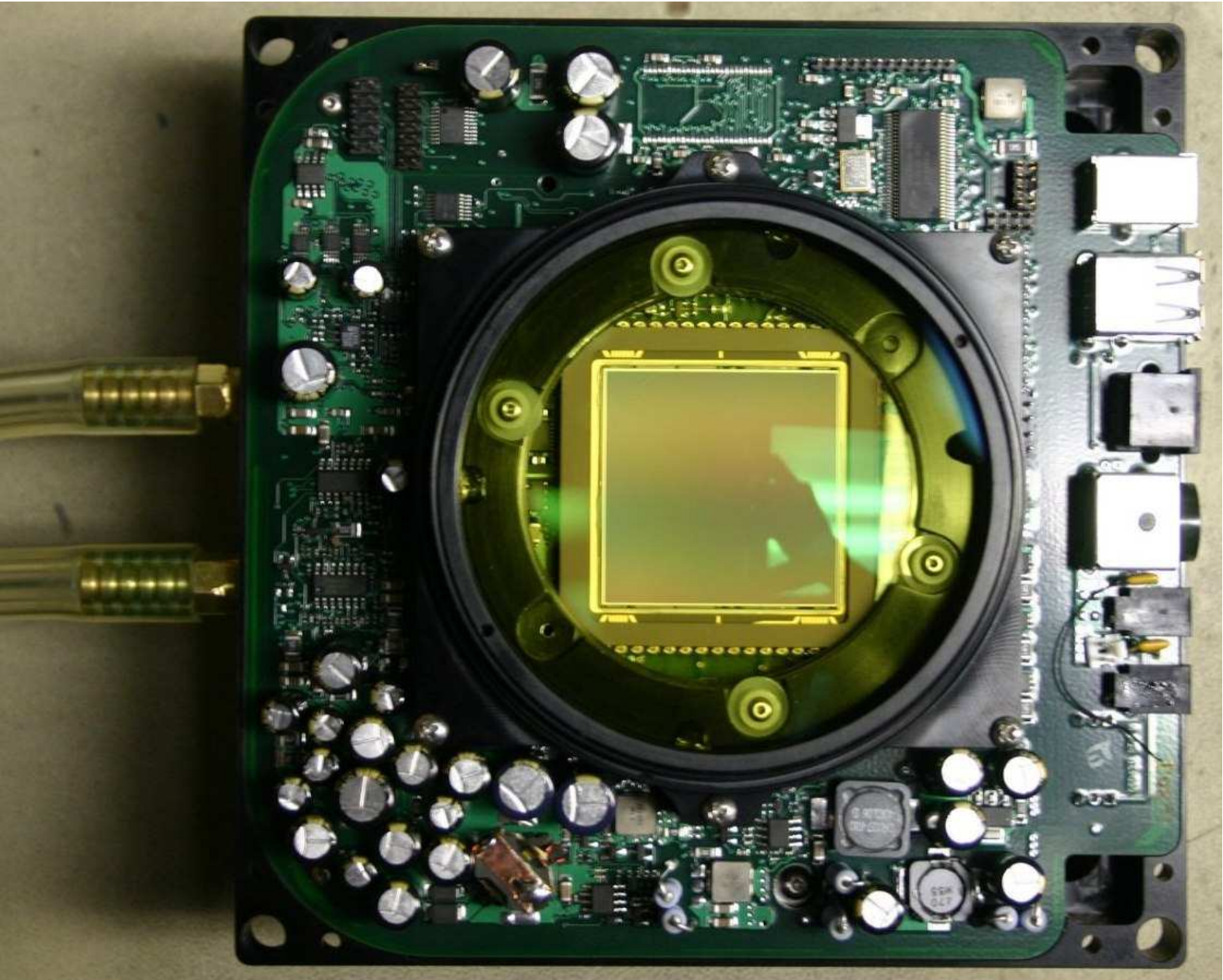}
\caption{Finger Lakes Instrumentation Inc. (FLI) ProLine 16803 CCD with field flattener lens, filter, and
glycol recirculating cooling system. Uniblitz CS90 90\,mm shutter and front of camera cover were removed.
Courtesy of FLI.}
\label{fig:ccd}
\end{figure}

Our custom design CCD was based on the production prototype PL16803 from Finger Lakes Instrumentation (FLI) Inc.
(New York, USA), as can be seen in Fig.~\ref{fig:ccd}. Its main specifications are: 4096x4096 9${\mu}$m-pixel
Kodak 16803 chip, 60\% QE at 550nm, 9-11\,e- readout noise, 1\,MHz and 8\,MHz readout speed at 16-bit digitization
rate, the camera electronics and sensor chambers are sealed with noble gas to keep the moisture out.

\placefigure{fig:ccd}

Aside from the above specifications, the CCD camera inside the TFRM tube had a number of requirements specific to
this project. For example, the field flattener lens and the filter must be placed inside the camera body. In
addition, conventional cooling by exhausting warm air from inside the tube is not an option. Therefore, a
recirculating liquid cooling system was implemented by FLI, Inc. Finally, compactness of the CCD camera dimensions
(6.2-inch${\times} 6.2$-inch) was also essential for fitting it into its spider housing. At the time of purchase,
the PL16803 was found to be the smallest large-format commercial CCD camera which met our requirements. A small
size was essential for our application to minimize the size of spider and CCD assembly, and therefore, the
obscuration over the image.

After taking stock and custom specifications into account, FLI, Inc. was found to be the commercial manufacturer
which best balances willingness of performing this custom design, quality of product and reasonable cost. 

Positioning the field flattener element within the camera body was one of the most critical aspects of the
project. The prescription of the refigured optics (see \S~\ref{opticsrefiguring}) required the flattening lens to
sit 0.65\,mm from the CCD sensor. Besides, the filter-sensor distance is such that the filter had to be placed
as the camera vacuum chamber window.

The procedure followed to assemble the corrective optics was supervised by one of us (MJM) and performed by FLI.
It was a complex sequence of precise measurements of the level of different parts of the camera and optics to be
assembled. From here, a few ${\mu}$m accurate positioning of the flattener lens and filter with respect to the CCD
sensor could be derived. The measurements were taken with a standard micrometer depth gauge while working on a
flat granite table. 

One of the 'extreme' characteristics of this optical system is the large angle (${\sim}60{\degr}$) light cone at
focus. The result is a required shutter diameter that increases very quickly with distance from the focal plane.
Unfortunately for us, most commercial shutters are simply too small to allow unobscured imaging with our camera
system. And, the requirement become even more stringent if the backfocal distance is to be kept within a
reasonable range. As a result, a large aperture shutter had to be considered. This device had to guarantee
reliable performance as well as mechanical and electrical stability, given the unattended nature of the TFRM. The
CS90 model from Uniblitz Inc. (New York, USA), with an aperture of 90\,mm, was selected. FLI assembled the shutter
internally to the PL16803 camera body, which had to be modified to accommodate the larger diameter of the shutter.
As part of the shutter integration process, FLI provided for both auxiliary and USB triggering of the shutter. 
Although the TFRM LAN latency could, in theory, provide a CCD image timestamp precision of ${\pm}0.1$\,ms, the
electrical and mechanical uncertainties of the CS90 shutter decrease the precision to about 100\,ms.

Although the CCD camera consumes a small amount of power (${\sim}40$\,W), one should note that it is stored inside
two cavities: its housing in the spider assembly and the tube. If conventional air cooling was employed,
turbulence within the tube would result - possibly decreasing image quality. As an alternative solution, our
camera was designed to accommodate liquid cooling. At the time of purchase, the FLI design was innovative in
commercial cameras and has shown a performance of typically +15{\degr} better cooling than the air option in the
same camera. In addition, the use of one cooling method does not preclude the other with this camera. The liquid
cooling option can accommodate a large amount of heat transfer from the camera to the outside environment. Should
the liquid cooling fail, air cooling can take its place. This, of course, would come at the expense of image
quality and cooling efficiency. As a coolant, we use a 25\% solution of propylene glycol in distilled water. This
is recirculated and maintained at a constant temperature by a ThermoCUBE$^{\rm TM}$, Solid State Cooling, Inc.
chiller. The ThermoCUBE$^{\rm TM}$ can be remotely controlled and monitored by the observatory control computer.

\subsubsection{Reinforced glass-fiber enclosure}

The TFRM and rest of observatory devices are installed in a new enclosure specially built for this purpose. The
enclosure has been designed, manufactured and assembled by GRPro Precision Manufacturing, Inc. GRPro has extensive
experience in astronomical enclosures, like the ones machined for SuperWASP (North and South) projects which were
source of inspiration for our design~\citep{don2007}.

\begin{figure}
\centering 
\includegraphics[angle=0,scale=0.38]{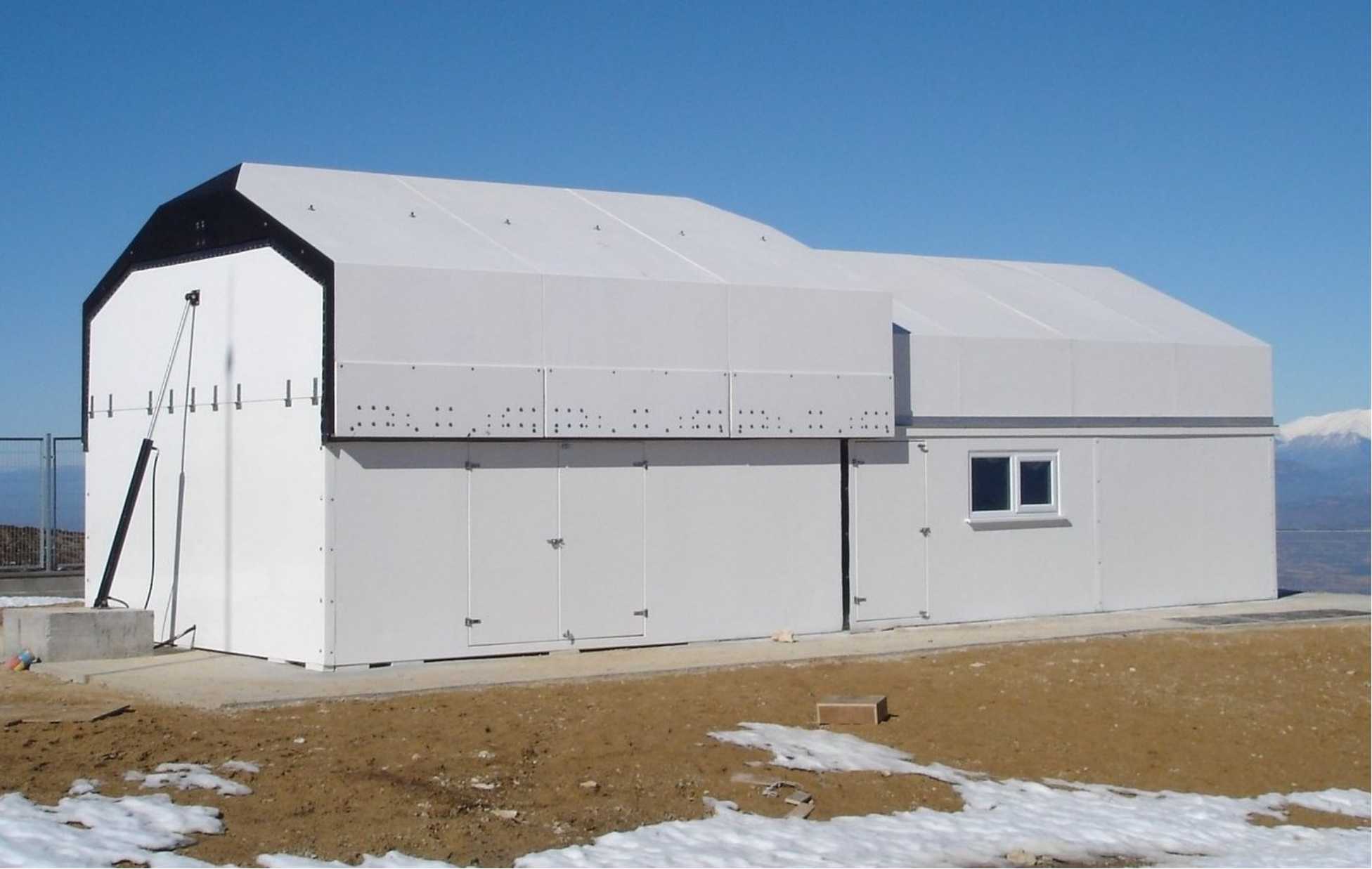}
\caption{TFRM reinforced glass-fiber enclosure installed at OAdM, with sliding roof and South gabling
wall.}
\label{fig:enclosure}
\end{figure}

The TFRM 12\,m x 5\,m x 4.5\,m reinforced glass-fiber enclosure is modular and allows a portable installation. It
has turned out to be robust in all kinds of adverse weather conditions with no mechanical failures. As seen in
Fig.~\ref{fig:enclosure}, the facility has a sliding roof which, when opened, it leaves the TFRM uncovered and
ready to observe. The rest of the building is dedicated to the control room. The South wall can be folded down by
90{\degr}, so that the TFRM can observe up to 13{\degr} elevations. 

\placefigure{fig:enclosure}

The roof and South wall are moved by means of an hydraulic pump which activates a mechanical chain, in the case of
the former, and an hydraulic arm in the case of the latter. In case of a power failure, backup $24$\,Vdc batteries
allow the system to close the enclosure. Start and end points of motions are monitored by means of mechanical
limit switches which stop the motor supply. 

A Vaisala MAWS100 meteorological station was installed at the communication tower of OAdM. This station is
composed by a WXT520 multisensor and two DRD11A precipitation sensors. This guarantees continuous monitoring
of enviromental conditions from the control software via TCP/IP protocol.

In order to introduce redundancy in meteorological recordings and gain safety in the overall TFRM operation, an
independent set of sensors (two DRD11A for precipitation, one for humidity, and one for daylight) were connected
to a watchdog system located at the enclosure. 

In contrast to control software which is running on a computer arquitecture, the watchdog system is a stand-alone
electronic device which runs off of backup battery power. In the case of a crash of the control computer
software, the watchdog will secure the facility by closing the enclosure roof and ram when sunrise or a weather
alert occur.

\subsubsection{Observatory control and scheduling software}

A state-of-the-art observatory control software based on a client-server architecture via a
Instrument-Neutral Distributed Interface (INDI) device communication protocol was created and developed by
one of us (ECD)~\citep{elwood2011}, who contributed to this refurbishment project as a consultant. All the
devices in the TFRM observatory communicate with their client and servers via INDI protocol, which is
designed to control a distributed network of devices in either remote or robotic fashion. All
communication uses TCP/IP sockets for reliable distributed operation.

Among other interesting features of INDI, we highlight the ability of clients to learn the properties of a
particular device at runtime using introspection. As a result, implementation of clients and devices are decoupled
which is crucial for code maintainance of both sides of a control software. Also, the protocol is XML-based for
passing parameters back and forth in a compact efficient format. Typical bandwidth requirements for monitoring and
control of all observatory functions (except camera images) are on the order of a few tens of kbps, so even simple
voice-grade modem connections are sufficient for routine remote operation.

Whenever INDI detects any of several conditions considered dangerous for further observations to continue it
issues a weather alert. This includes excessive wind speed, humidity, detection of rain, hail and snow, high
levels of electrostatic atmospheric activity, and low UPS battery power. When an alert is issued the system
automatically closes the enclosure roof and ram. The INDI configuration contains a parameter that allows adjusting
the length of time an alert will remain in effect after any or all causal factors have returned to normal.

INDI drivers were developed for all devices at the observatory which need active command or record. Drivers are
written in ANSI C for the Linux operating system. Within each driver is the code that implements the desired
functionality for one, and only one, INDI device. Some drivers only provide services, such as target prediction.
Other drivers control hardware. Drivers may also communicate with other drivers. The INDI architecture places no
restrictions on what a driver can do. The only requirement is that it responds to INDI messages that arrive on its
stdin stream for its device and that it generates valid INDI messages from its device on its stdout stream.

Clients, like drivers, may do anything they wish so long as they communicate valid INDI messages over the
socket with which they connect to an indiserver. Otherwise clients can be GUIs, command line programs,
daemons or other process roles and may be written in any desired language. Java language was chosen for
the development of INDI GUIs clients, so that maximum portability and consistency across platforms
(Linux under KDE o Gnome, Windows and Mac OS) was assured. 

\begin{figure*}
\centering 
\includegraphics[angle=0,scale=0.79]{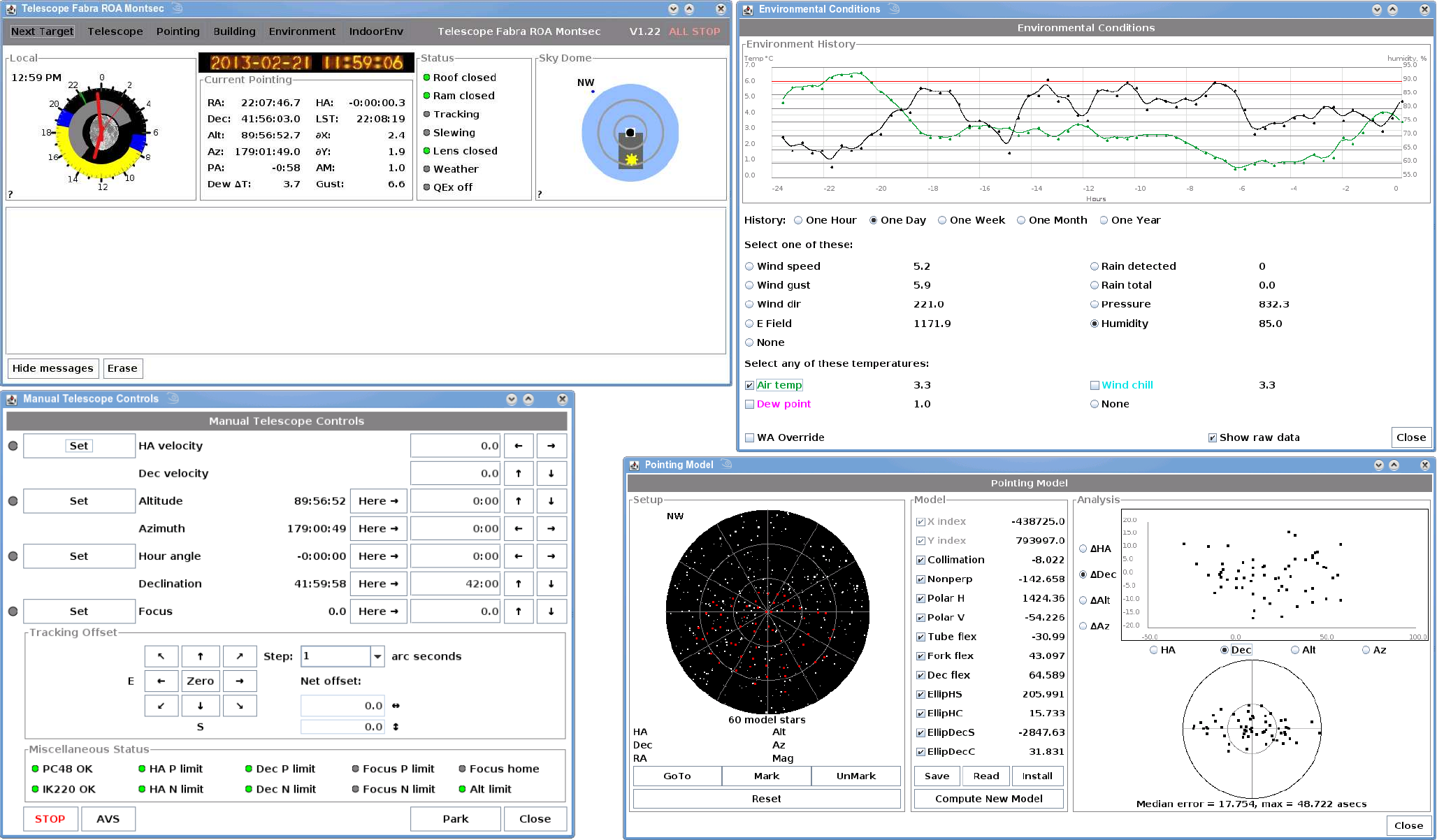}
\caption{Four windows of I-INDI for TFRM remote control. {\bf Upper left:} Main window which summarizes
the status of critical devices and circumstances. {\bf Upper right:} Environmental variables, both
instantaneous values and their plot versus time. {\bf Lower left:} Manual telescope control. {\bf Lower
right:} Pointing model with up to 13 coefficients and plot of reference stars and errors.}
\label{fig:indi}
\end{figure*}

A short description of INDI clients follows:
\begin{itemize}

\item[1.] I-INDI (stands for Interactive-INDI) provides remote command and monitoring capability for all
observatory systems except the CCD camera. See in Fig.~\ref{fig:indi} snapshots of some of the I-INDI
windows for the global status of most important devices, and control of environmental variables, telescope
pointing and pointing model.

\item[2.] S-INDI (stands for Sheduled-INDI) allows to dispatch robotic operations, whose observing blocks
were previously written in XML format. Using S-INDI you define the INDI commands you want to execute,
define the target and any additional constraints for the observation, then the S-INDI device driver will
decide the best time to perform the request. Many requests may be pending simultaneously and the S-INDI
driver will always attempt to perform each of them at the best possible time.

\item[3.] CCD-INDI commands the CCD camera in a remote fashion. It can also read and write FITS files from
and to disk. It is intended only as a basic camera control and image display tool. It is not intended to
compete with very elaborate control and processing tools.

\item[4.] ANSI C language was chosen for the development of simple command line clients. These were
conceived for the purpose of implementing complex environmental conditions decisions via high-level
scriptable languages (Perl, Python or bash) that can be scheduled on the crontab of the observatory
control computer.

\end{itemize}

\placefigure{fig:indi}

\subsection{Observing site}
\label{site}

The TFRM was installed at the Observatori Astron\`omic del Montsec (OAdM), in the Catalonian Pre-Pyrenees, whose
WGS84 coordinates are: ${\phi} = 42{\fdg}0516$ N, ${\lambda} = 0{\fdg}7293$ E, and h = 1570\,m HMSL. To date, the
OAdM is pioneered by the Consorci del Montsec, an institution run by the Catalonian Government. The observatory is
located at the Montsec d'Ares mountain, 50\,km South of the central Pyrenees, in the province of Lleida (Spain).
The site was chosen after a site-testing campaign. The OAdM also hosts the 0.8\,m Joan Or\'o Telescope, named in
honour of this famous Catalonian researcher.

The installation of the TFRM at OAdM resulted in a number of infrastructure upgrades to the facility as a whole:
stable power line, a 100\,Mbps Internet access via fiber optics cable, and enhanced security fence.

\subsection{Reproducibility of the refurbishment process}
\label{reproducibility}

Including the TFRM, four BNCs have already succesfully refurbished, and another one (ARIES in India) is in
process. 

This demonstrates that the combination of the current synergies between information technologies, devices control
electronics, and control software advances enable the upgrading of this kind of telescopes into a facility with
unique specifications and great scientific potential. From the originally manufactured 21 BNCs, there are still a
good number that are inactive but in good shape, which could benefit from a refurbishment project like ours.

Furthermore, in the case of TFRM, a number of steps of the refurbishment project were accomplished by
purchasing commercial components, which involve less cost and little posterior engineering assembling
work.

\section{ONGOING SURVEYING PROGRAMS}
\label{surveying_programs}

\subsection{Transiting exoplanets}

The large FoV of the TFRM, together with its moderate aperture and robotic nature, allows for the efficient
detection of exoplanets by means of transit measurements with high signal-to-noise ratio in the appropiate
magnitude range. The suitability of such an instrument for exoplanet research was confirmed earlier by the APT
during their UNSW Extrasolar Planet Search during the period of 2004-2007~\citep{2008MNRAS.385.1749C}. The
subsequent catalogue that they produced shows that refurbished BNCs can accomplish millimagnitude photometry at
least up to V$\sim$14\,mag.

\begin{figure*}
\centering 
\includegraphics[angle=0,scale=0.92]{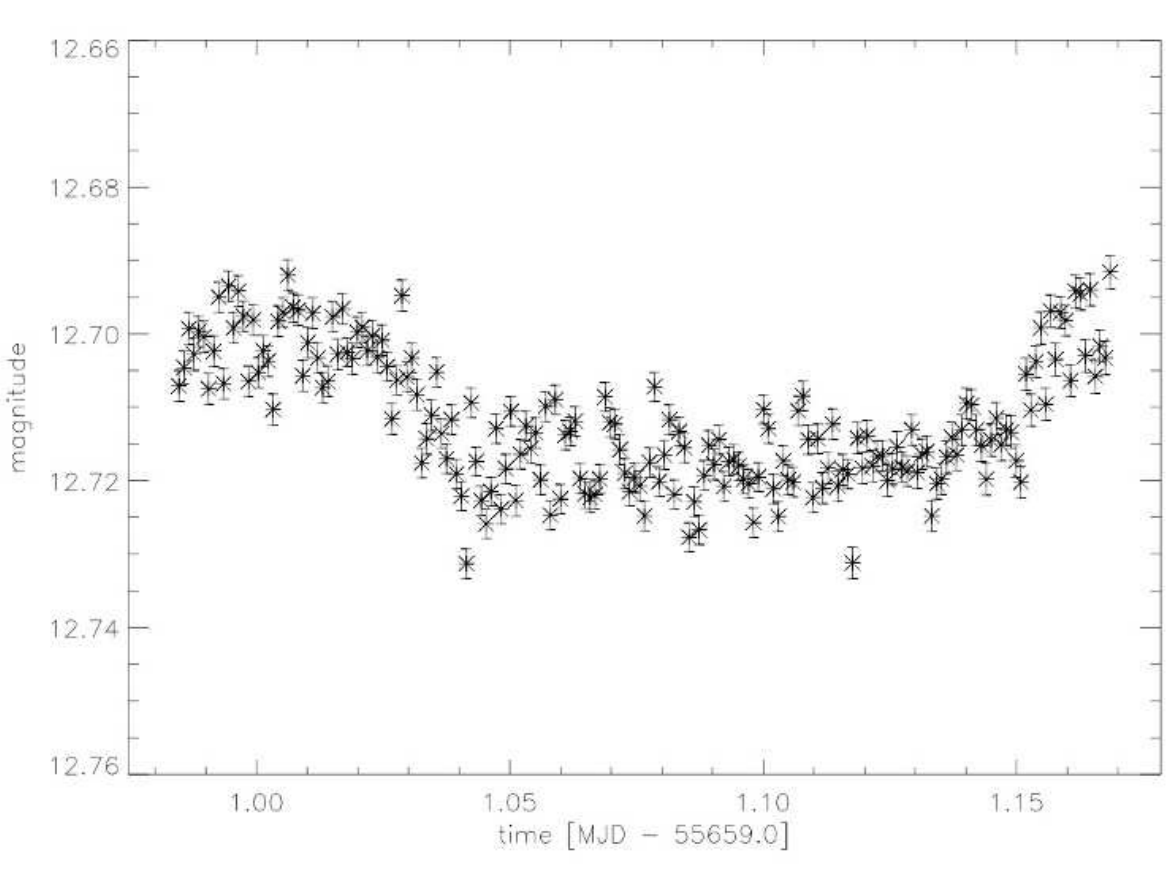}
\caption{WASP-37b transit lightcurve observed by TFRM on Apr 8th 2011. Note that the left and right
baseline is not completely spanned because that corresponds to a single-night observation. Photometric
precision is 4.3\,mmag.}
\label{fig:wasp37b}
\end{figure*}

In order to confirm the APT performance, the TFRM observed a predicted transit of WASP-37b, a known
exoplanet, in a completely unsupervised robotic mode on 8 Apr 2011. The first transit-like signatures of WASP-37b
were detected by SuperWASP-N survey (La Palma) between March and June in 2008 and 2009, and by SuperWASP-S survey
(South Africa) during 2008 June to July and 2009 March to July. The transit lightcurve spanned about 4.5 hours
(see Fig.~\ref{fig:wasp37b}). Transit analysis was carried out by Holger Voss using the reduction software
described in~\citet{voss2006}. The photometric performance shown by the TFRM was outstanding: differential
photometric precision of 4.3\,mmag for WASP-37b (V${\sim}12.7$\,mag), and 3\,mmag for stars of similar magnitude.
Aside from the excellent precision, what is most relevant is that, if WASP-37b were unknown, TFRM would have
detected it as an exoplanet candidate on the very first night of observation, i.e., like a real-time detection
without the need of further phase-folded data points of posterior nights. 

\placefigure{fig:wasp37b}

Other known exoplanets transits were observed by TFRM, all of them in the 12\,mag$<$V$<$14.5\,mag range,
with similar photometric precisions in all cases.

\citet{2009IAUS..253...37I} proposed an interesting alternative observational approach which has been executed by
the MEarth project. In order to maximize the probability of detection of rocky super-Earths in the Habitable Zone
(HZ), MEarth is photometrically monitoring a sample of ${\sim}2000$ M-type stars, which have been pre-selected.
MEarth operates 8 telescopes f/9 Ritchey-Chretien with a field of view of 25{\arcmin}x25{\arcmin} each. Due to
this limited FoV (0.17 square degrees), this project can only monitor a single star per telescope at a time.
Despite this limitation on the efficiency of the survey, only in 3-4 years of full operation, it has been able to
detect the first super-Earth (GJ1214b) with this new pre-selected strategy survey~\citep{2009Natur.462..891C}.

As mentioned in~\citet{2010ASPC..430..428F}, the 19.4 square degrees TFRM FoV is the most remarkable
feature of this telescope. This, combined with the fact that a 30-second exposure typically contains
${\sim}20.000$ stars with SNR$>$5 (V$<$15.5\,mag) and a photometric precision better than 10\,mmag
(3-4\,mmag typically for V down to 13-13.5\,mag), means that the telescope has a significant probability
in detecting new exoplanets by transit technique.

Since December 2011, and in collaboration with the team of Dr.\,Ignasi Ribas (ICE-CSIC), the TFRM began to survey
a pre-selected series of fields, with an input catalog similar to
MEarth's~\citep{1995AJ....110.1838R,1996AJ....112.2799H,2011AJ....142..138L}, in search of super-Earths around
M-type stars. The survey was called TFRM-PSES (TFRM-Preselected Super-Earths Survey). TFRM-PSES monitors a number
of M0 to M5-type catalogued targets comprised in several fields with sufficient frequency each night, and in the
range of 9.0\,mag$<$V$<$15.5\,mag. M targets per field distribution spans from 6 to 16, with a global median value
of 8. However, up to 23 out of more 60 fields contain more than 13 M targets: typically 14 and 15, and 16 in one
case. Note this is the main difference between MEarth and TFRM-PSES: on one hand, while in MEarth each single
telescope monitors one star per CCD field, TFRM-PSES captures approximately 8 times as many stars per field which,
therefore, increases survey efficiency. On the other hand, the higher number of telescopes in MEarth compensates
the former said. Finally, that TFRM-PSES magnitude limit of V=15.5\,mag could be increased, but then the frequency
of measurements would be less, which would penalize efficiency when recording possible transits.

\begin{figure*}
\centering 
\includegraphics[angle=0,scale=0.80]{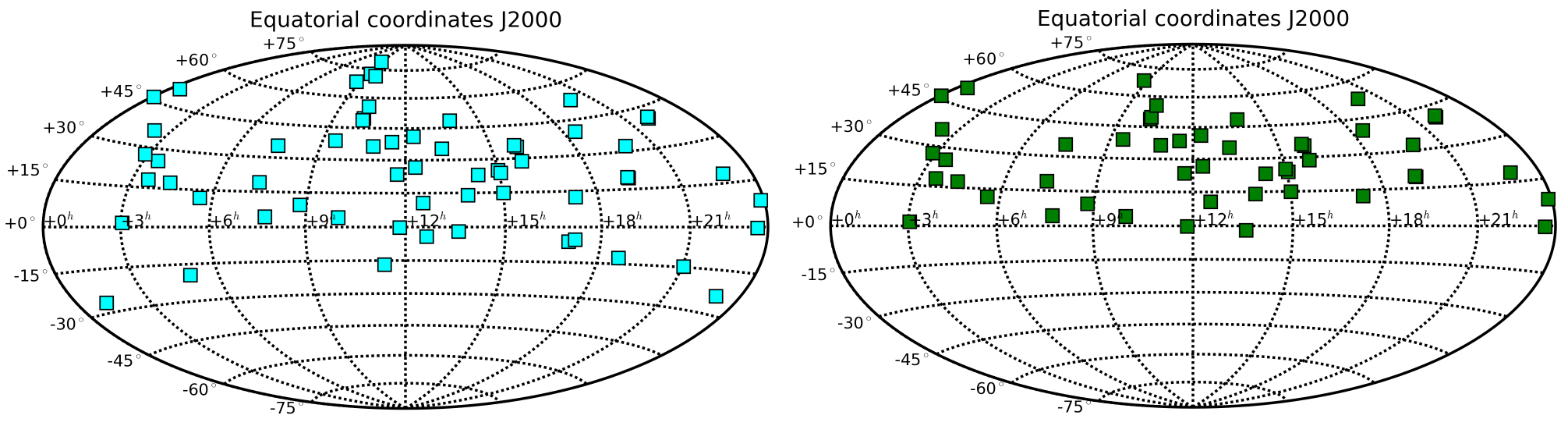}
\caption{{\bf Left:} Catalogued fields with M dwarfs to be observed. {\bf Right:} Fields with M dwarfs
already observed.}
\label{fig:pses}
\end{figure*}

In particular, as seen in Fig.~\ref{fig:pses} the coverage of the TFRM-PSES survey to 5 Apr 2013 was such
that 48 of the 60 catalogued fields were observed at least once. The median number of epochs is 12, and the
total number of covered fields per night including repetitions is 635.

\placefigure{fig:pses}

Preliminary results with respect to the photometric precision and exoplanets detection probability of TFRM-PSES
survey were presented in~\citet{CS17}. This study showed that photometric precision down to 5\,mmag is achieved
in the range of 11.0\,mag$<$V$<$14.0\,mag. A more in-depth study of the TFRM-PSES performance and subsequent
detections is in process of publication.

A by-product result of TFRM-PSES survey is the detection of new variables stars. Although we cannot provide
detailed statistics of detection, a good example is the WASP-37b transit observation formerly presented: in 4.5
hours of photometric measurements of all the objects in the $4{\fdg}4{\times}4{\fdg}4$ FoV, ten new variable stars
of different types and magnitudes were detected.

\subsection{Space debris}

Among other TFRM capabilities, its $4{\fdg}4{\times}4{\fdg}4$ FoV, the telescope tracking at arbitrary $\alpha$
and $\delta$ rates, and the CCD shutter commanding at will during the exposure are extremely useful for the
participation in Space Situational Awareness / Space Surveillance and Tracking (SSA/SST) observational programs.

\begin{figure*}
\centering 
\includegraphics[angle=0,scale=0.75]{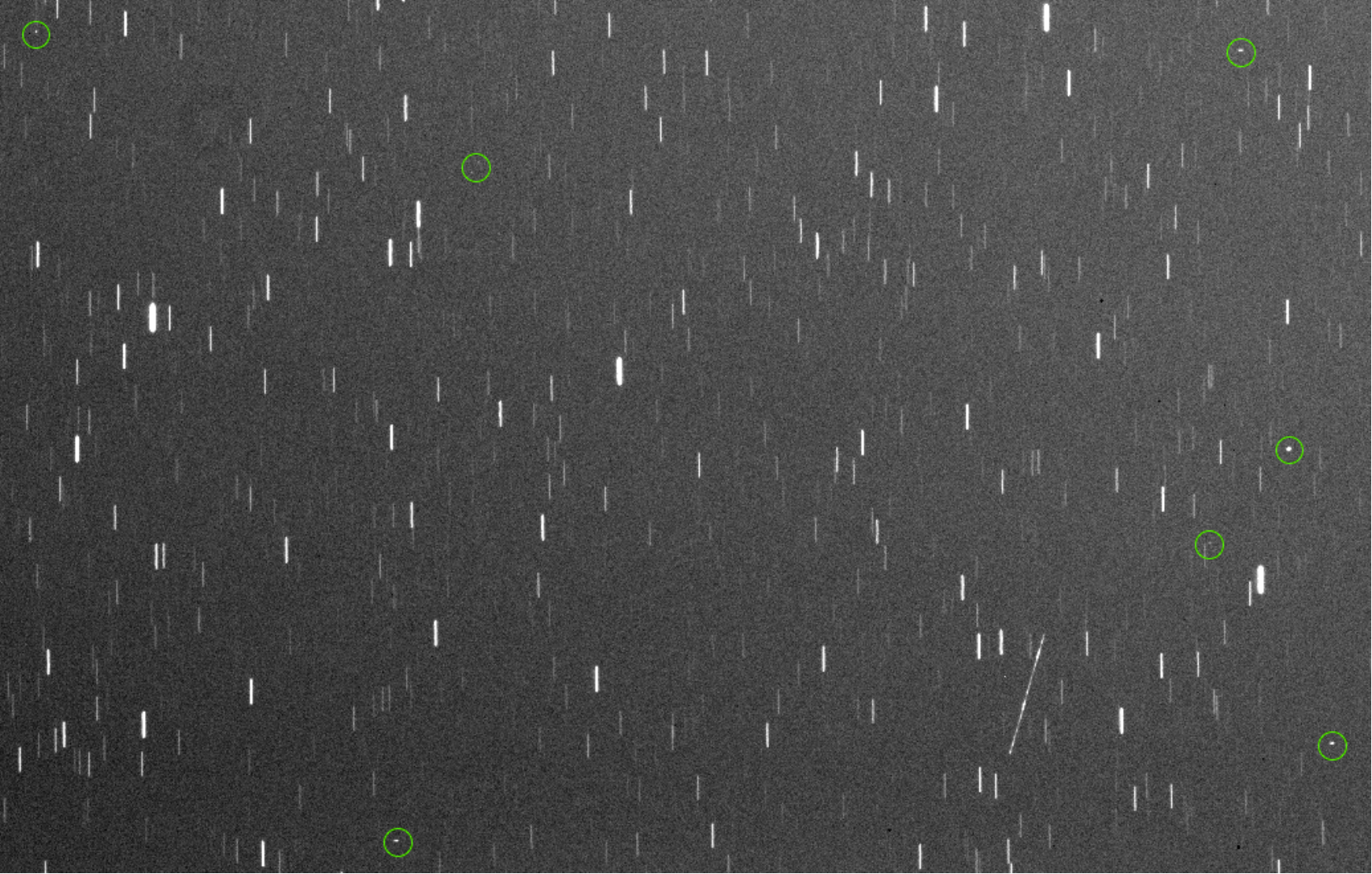}
\caption{A 10s exposure taken with the TFRM where 8 GEO objects (two inside the same green circle forming a
constellation) and another object in a lower orbit are easily identifiable among the trailed background stars.
Only half of the TFRM FoV is shown. The orientation of the image is North-leftwards and East-upwards.}
\label{fig:eightGEOs}
\end{figure*}

The TFRM's large FoV is suitable to survey the entire visible geostationary belt from its location. In fact,
with TFRM we can cover twice our entire visible geostationary belt in a 12 hours night. The best method to track
and detect objects close to the geostationary orbit (GEO) is with the telescope stopped, i.e. in a Earth-fixed
Reference System. So the background stars will appear as trails with length proportional to the exposure time and
the objects in the GEO belt will appear as quasi-point-like sources. A good example of the TFRM's detection
capability in a single image is the Fig.~\ref{fig:eightGEOs}, where in half FoV there are easily identifiable 8
GEO objects (two inside the same circle forming a constellation) among the trailed background stars.

\placefigure{fig:eightGEOs}

Furthermore, the telescope's capability of tracking at arbitrary $\alpha$ and $\delta$ rates jointly with the
software control, permits the tracking of objects in any kind of orbit even Low Earth Orbits (LEO), by simply
entering its Two Line Elements (TLEs) in the INDI target property.

Commanding the CCD shutter at will during the exposure could be very useful for surveying the sky looking for
objects in any kind of orbit. This observing approach allows to cut the object trails while the sidereal tracked
exposures are timestamped. Nevertheless, this method has not been tested yet.

The TFRM's collaboration in the SSA/SST international effort develops in two different projects: the
European Space Agency (ESA) program and the International Scientific Optical Network (ISON) survey. 

TFRM is one of the Spanish assets that is involved in the ESA SSA/SST Preparatory Programme (2009-2012).
Telescopes and radars from other European countries also participate in this project. 

During 2011 TFRM took part in the third ESA CO-VI 7-day observational campaign. This was an experimental satellite
tracking campaign using European facilities, aimed to determine how accurately existing telescopes can work
together to track objects in geosynchronous orbits. The satellite positions of every asset were submitted to the
coordinating office at European Awareness Research Laboratory for Space (Early-Space), which reported the global
results of the campaign~\citep{caroline2011}. Systematic observations of different GEO satellites were conducted
by TFRM to determine 1137 satellite angular positions, and partial TFRM results were
presented~\citep{2011arXiv1109.5918M}. We estimate our astrometric precision in the GEO satellites angular
coordinates to be below $0{\farcs}5$ in both coordinates. 

\begin{figure*}
\centering 
\includegraphics[angle=0,scale=1.0]{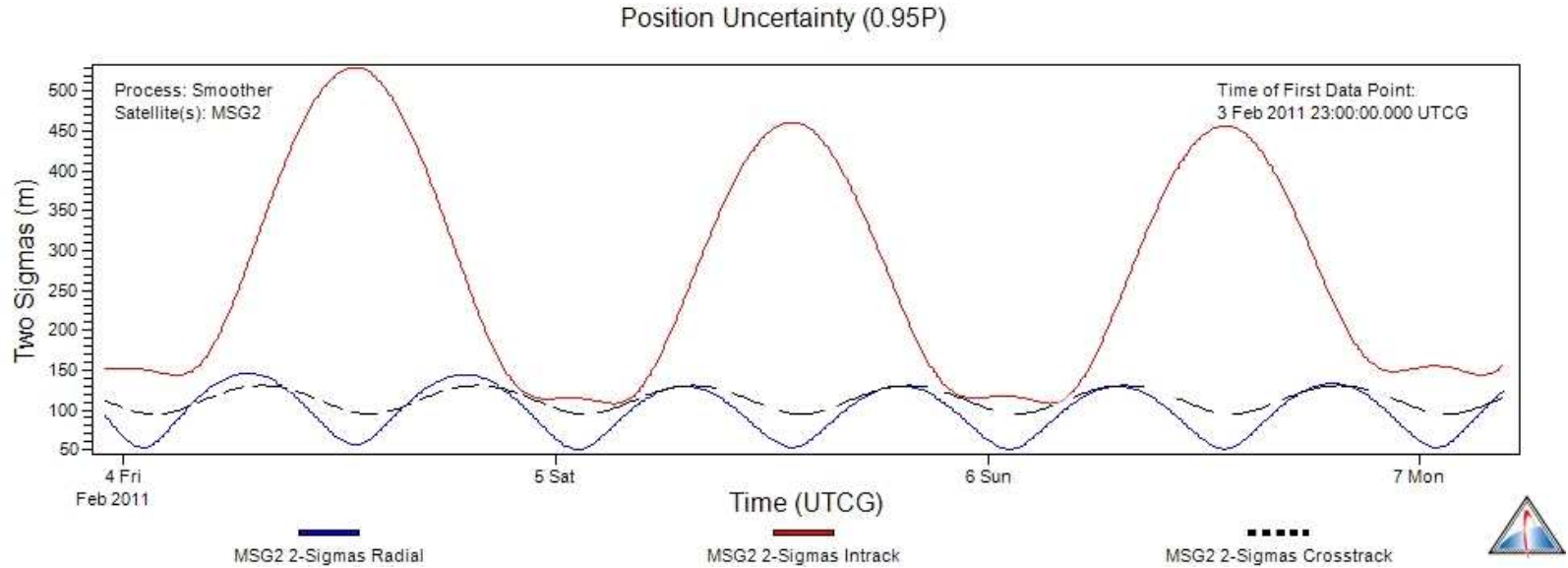}
\caption{Uncertainties in the position of the MSG2 satellite, at 2 sigmas (95\%) and in RIC coordinates. 
2-Sigmas Intrack in red solid line. 2-Sigmas Radial in blue solid line. 2-Sigmas Crosstrack in black dashed line. 
It is clearly noticeable the characteristic increasing of the Intrack uncertainty during daytime.}
\label{fig:odtk}
\end{figure*}

In order to test our data quality, orbit determination from the angular measurements was carried out using the
Orbit Determination Tool Kit (ODTK) software package, from Analytical Graphics, Inc. (AGI). As an example, in
Fig.~\ref{fig:odtk}, we show 2-sigma (95\%) uncertainties obtained over the MSG2 satellite, with 175 angular
measurements along 4 nights in which the satellite was not maneuvered. The mean uncertainties in the classical
elements, i.e., semiaxis, eccentricity and inclination, are of the order of 12\,m, $1.8{\cdot}10^{-6}$ and
$1{\fdg}5{\cdot}10^{-4}$, respectively. It is worth mentioning that during this campaign the TFRM was still in
commissioning period and the GEO objects reduction process was performed using a non-automated and non-optimized
software based on SExtractor. Nowadays we can take advantage of the advanced and fully automated reduction
software APEX-II developed by Vladimir Kouprianov (Pulkovo Observatory).

\placefigure{fig:odtk}

At the time of writing the TFRM is about to participate in the imminent upcoming ESA campaing \textquotedblleft
CO-VI Optical Observations for Space Surveillance and Tracking Test and Validations\textquotedblright.

The International Scientific Optical Network (ISON) is a civilian non-governmental project devoted to space debris
research and space situation awareness. TFRM is collaborating with ISON in its sistematic survey of the GEO
Protected Zone since 2011~\citep{agapov2011}. Positional measurements are derived using advanced trailed image
reduction techniques included in APEX-II sofware~\citep{2010SoSyR..44...68D}. As a result of this collaboration,
the TFRM is one of the sensors that contributes to the completeness of the objects without Two-Line-Element data
of ESA's DISCOS database, as stated at the last \textquotedblleft Classification of Geosynchronous Objects
Report\textquotedblright~issued by ESA~\citep{floher2012}.

Currently TFRM is observing routinely and can detect an average of 400 GEO objects tracks per night with an
accuracy better than $0{\farcs}5$ in both coordinates and a limit magnitude of 16\,mag. Furthermore, the TFRM team
is in the process of improving the limit of detection towards fainter GEO objects~\citep{Fors2010}.
Typically in a 12 hour night the TFRM is measuring around 2800 positions of 320 different objects.

\begin{figure}
\centering 
\includegraphics[angle=0,scale=0.71]{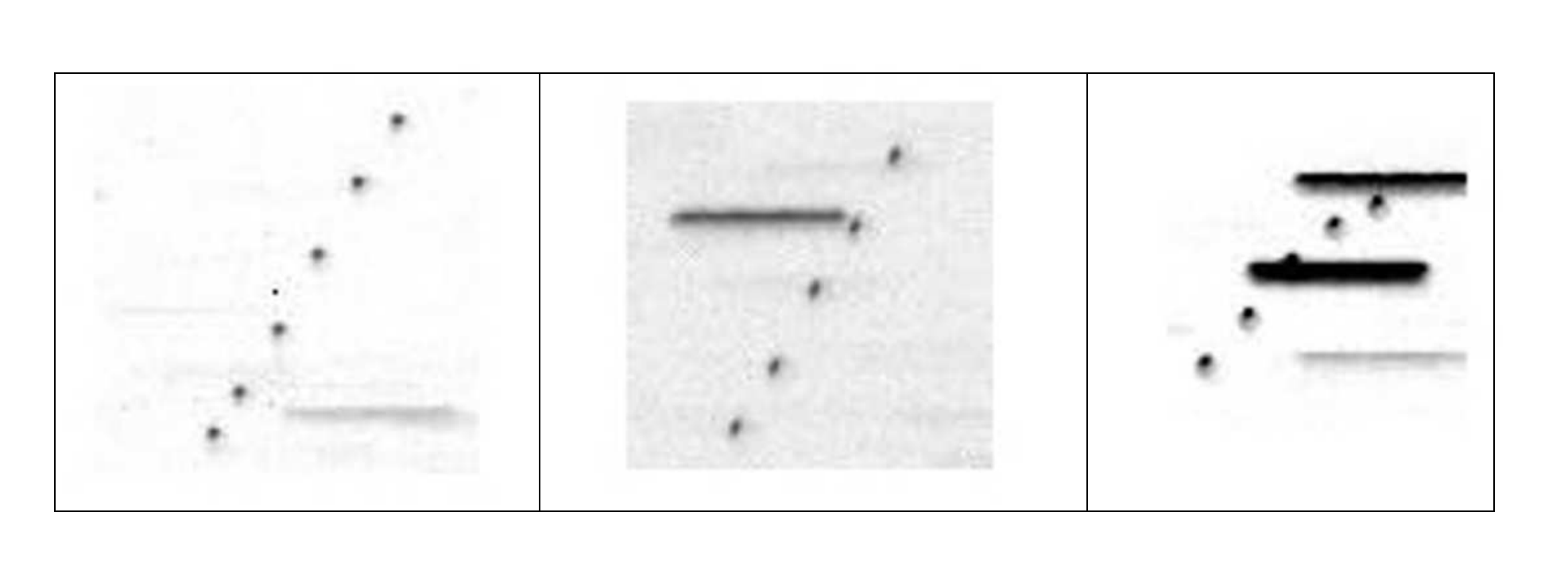}
\caption{Three tracks of MSG-3 satellite (point-like features) imaged by TFRM, while the satellite orbit was
maneuvering East at 3{\degr} per hour.}
\label{fig:msg3}
\end{figure}

A good example of the TFRM's capabilities in the SST field was the early detection after the MSG-3 (Meteosat 10)
satellite launch. This GEO satellite was on its way after lifting off on an Ariane 5 at 21:36 UTC on Thursday, 5
July from Europe's Spaceport at the Guiana Space Centre in Kourou, French Guiana. The MSG-3 was first detected by
TFRM on the night of 12 July, during our routine collaboration in the ISON geosynchronous space survey. Three
tracks (see Fig.~\ref{fig:msg3}) were detected over the night with the automatic GEO objects detection software
APEX-II. With additional follow-up observations from other telescopes of ISON network, an initial orbit
determination was performed by ISON before the satellite TLEs were published, and the results showed that the
satellite was indeed the MSG-3, which was drifting East at 3{\degr} per hour rate. Hence, it was caught
maneuvering to its final 0{\degr} longitude expected geostationary slot.

\placefigure{fig:msg3}

\subsection{Other observational programs}

A number of other observational programs can benefit from TFRM specifications.

One is the collaboration in the Space Situational Awareness/NEO-segment (SSA/NEO). The TFRM will also be
one of the assets involved in the ESA SSA/NEOs segment. NEOs are small solar system bodies whose orbits
bring them close to Earth and which represent a potential threat to the Earth. The NEO segment of the
European SSA System will perform observations of NEOs, predict their orbital evolution and impact risk,
store observational and calculated data, issue NEO information, news releases and impact warnings and
support NEO mitigation measures. Concerning NEO observations, ESA is planning to scan every night the
complete visible sky with the aim to detect objects which are only visible when they are close to Earth.

In the same context of SSA/NEO, the TFRM has the capability to contribute significantly to the international effort of
surveying and monitoring the population of NEOs: the observations of NEOs by TFRM includes imaging asteroids at low solar
elongation (an area usually poorly searched) in collaboration with the NESS project, led by Dr. Alan R. Hildebrand (University
of Calgary) that will use the NEOSSat microsatellite to continuously search in this near-Sun region.

The similarities between the SSA/NEO and TFRM-PSES survey strategies make that one program can partially benefit
from other's data. Furthermore, other survey programs related with the search of Solar System objects, like main
belt asteroids, comets, KBOs, and TNOs are also partially compatible.

Another program which was initiated is the optical monitoring of $\gamma$-ray binaries. The final aim is to study
how the relativistic wind of the young non-acreating pulsar affects the circumstellar envelope in $\gamma$-ray
binaries through optical photometric variability. Preliminary observations in the case of \object{HD~215227} were
presented in~\citet{2012arXiv1210.1151P}.

Finally, other alert programs, such as GRBs, SNs, novae, blazars, and other transients in general can be
allocated with the proper observational strategy.

\section{DISCUSSION}
\label{discussion}

\begin{figure*}
\centering 
\includegraphics[angle=0,scale=0.36]{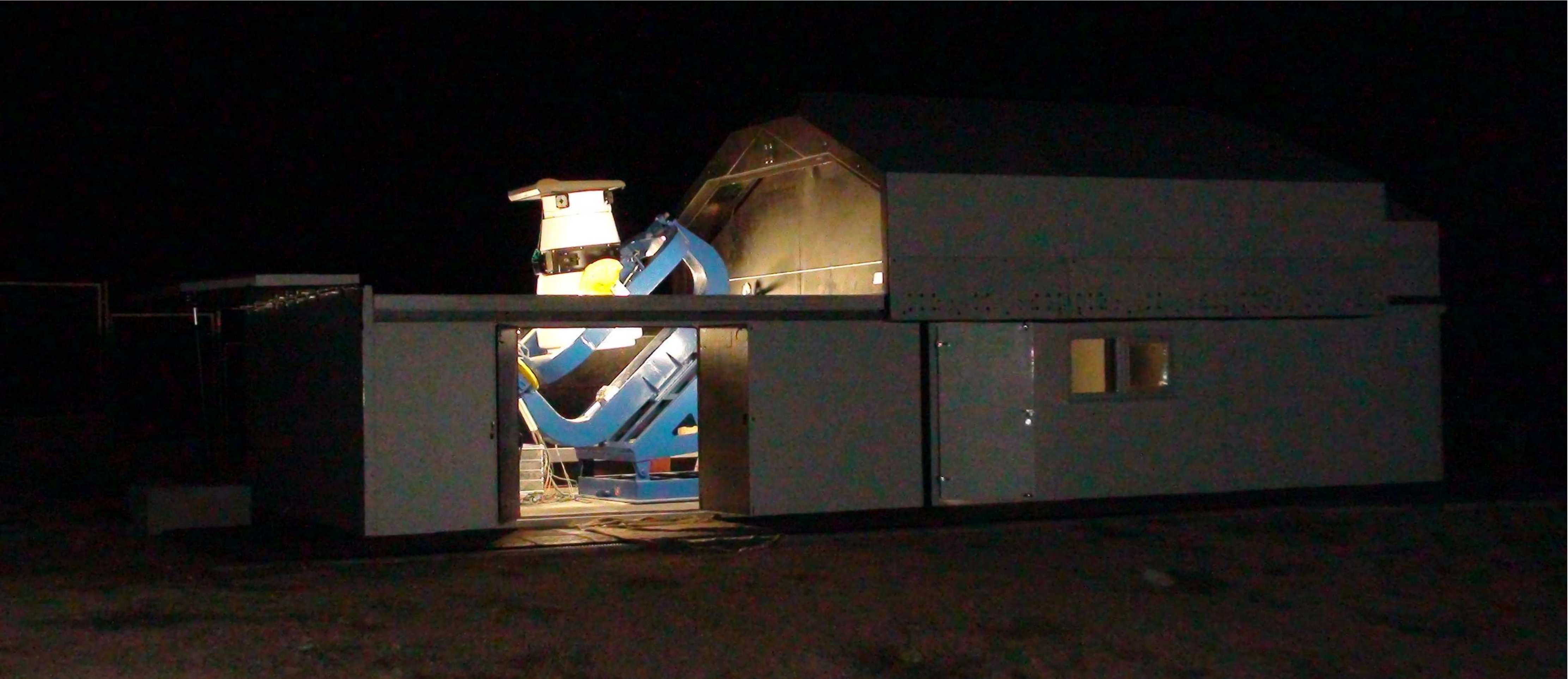}
\caption{Night view of the TFRM observatory.}
\label{fig:nightviewTFRM}
\end{figure*}

A Baker-Nunn camera has been refurbished to operate with a large-format commercial CCD camera in remote
and robotic modes. A night view of the TFRM ready to observe is shown in Fig.~\ref{fig:nightviewTFRM}.

\placefigure{fig:nightviewTFRM}

The refurbishment project included several steps, such as modification of the mount into a motorized equatorial
type, manufacture and installation of a new precise spider and focus system, optics refiguring for flattening the
CCD focal plane, customization of CCD camera, new reinforced glass-fiber enclosure with sliding roof and folding
down South wall, and new observatory control and scheduling software among others. Most of these steps were 
executed by the authors. When required, specialiazed external personnel was hired (spider manufacture, CCD
customization, reinforced glass-fiber enclosure). The rest of work was carried out by purchasing commercial
components and assembling them with little engineering time.

The performance of the TFRM was shown by two different survey-type programs: millimagnitude precise photometry of
exoplanets transits, and geostationary debris in the context of Space Situational Awareness / Space Surveillance
and Tracking (SSA/SST) programs. 

All-in-all, the acquired {\it know-how} and the research return of the new refurbished facility fully
justifies the cost involved in the project, which is affordable even for small research institutions or
Universities.

Furthermore, a number of other BNCs are still inactive and stored in good conditions, ready to be
refurbished. With the usual decrease of cost when replicating a project, the TFRM refurbishment project
could be applied to such BNCs, and enable in a short term basis (1-2 years) each of these telescopes in a
scientific useful facility.

\begin{acknowledgements}

This effort was supported by the Ministerio de Ciencia e Innovaci\'on of Spain (AyA2008-01225 and others), and by
Departament d'Universitats, Recerca i Societat de la Informaci\'o of the Catalonian Goverment (several grants). We
acknowledge Alan R. Hildebrand and Robert D. Cardinal for sharing their RAO BNC experience with us, which was highly
beneficial for the proper development of TFRM. This project has benefited from the wise advise of Ignasi Ribas, Don
Pollacco and Juan Carlos Morales, as well as from the data analysis expertise of Holger Voss and Vladimir Kouprianov.
Authors also acknowledge Daniel del Ser, and Albert Rosich and Estela Mart\'{\i}n-Badosa for their help in several figures
generation with Python and ZEMAX, respectively. Orbit Determination Tool Kit (ODTK) Analytical Graphics, Inc. (AGI) is a
licensed software used by ROA in collaboration with the Instituto Nacional de T\'ecnica Aeroespacial (INTA). 

\end{acknowledgements}

\clearpage

\end{document}